\documentstyle[12pt]{article}
\def\theequation{\thesection.\arabic{equation}}
\newcommand{\be}{\begin{equation}}
\newcommand{\ee}{\end{equation}}
\newcommand{\ba}{\begin{array}}
\newcommand{\ea}{\end{array}}
\newcommand{\bea}{\begin{eqnarray}}
\newcommand{\eea}{\end{eqnarray}}

\def\vtheta{\vartheta}
\def\bbbr{{\rm I\!R}} %real numbers
\def\bbbz{{\mathchoice {\hbox{$\sf\textstyle Z\kern-0.4em Z$}}
{\hbox{$\sf\textstyle Z\kern-0.4em Z$}}
{\hbox{$\sf\scriptstyle Z\kern-0.3em Z$}}
{\hbox{$\sf\scriptscriptstyle Z\kern-0.2em Z$}}}}
\def\cn{\hbox{cn}}
\def\dn{\hbox{dn}}
\def\sn{\hbox{sn}}

\def\i{\hbox{i}}

\def\dfrac#1#2{\frac{\displaystyle #1}{\displaystyle #2}}

\def\spaceTT{\vphantom{\bigg)}}

\chardef\ii="10
\def\vphi{\varphi}
\def\vrho{\varrho}
\def\vtheta{\vartheta}
\def\rphi{\vphi\in[0,2\pi)}

\def\SCPZ{{\xi^2+\eta^2}}

\def\onehalf{{\textstyle \frac12}}

\def\dfrac#1#2{\frac{\displaystyle #1}{\displaystyle #2}}

\def\bbbc{{\mathchoice {\setbox0=\hbox{\rm C}\hbox{\hbox
to0pt{\kern0.4\wd0\vrule height0.9\ht0\hss}\box0}}
{\setbox0=\hbox{$\textstyle\hbox{\rm C}$}\hbox{\hbox
to0pt{\kern0.4\wd0\vrule height0.9\ht0\hss}\box0}}
{\setbox0=\hbox{$\scriptstyle\hbox{\rm C}$}\hbox{\hbox
to0pt{\kern0.4\wd0\vrule height0.9\ht0\hss}\box0}}
{\setbox0=\hbox{$\scriptscriptstyle\hbox{\rm C}$}\hbox{\hbox
to0pt{\kern0.4\wd0\vrule height0.9\ht0\hss}\box0}}}}

\newcommand{\fcaption}[1]{
        \refstepcounter{figure}
        \setbox\@tempboxa = \hbox{\footnotesize Fig.~\thefigure. #1}
        \ifdim \wd\@tempboxa > 5in
           {\begin{center}
        \parbox{5in}{\footnotesize\smalllineskip Fig.~\thefigure. #1}
            \end{center}}
        \else
             {\begin{center}
             {\footnotesize Fig.~\thefigure. #1}
              \end{center}}
        \fi}
\newcommand{\tcaption}[1]{
        \refstepcounter{table}
        \setbox\@tempboxa = \hbox{\footnotesize Table~\thetable. #1}
        \ifdim \wd\@tempboxa > 5in
           {\begin{center}
        \parbox{5in}{\footnotesize\smalllineskip Table~\thetable. #1}
            \end{center}}
        \else
             {\begin{center}
             {\footnotesize Table~\thetable. #1}
              \end{center}}
        \fi}
\begin{document}

\begin{center}
{\Huge\bf
Separation of Variables

\vspace{0.3cm}
and Lie Algebra Contractions.

\vspace{0.3cm}
Applications to Special Functions}

\vspace{1cm}
{\Large\bf George Pogosyan$^{1,2,3}$, Alexey Sissakian$^{2}$,

\vspace{0.3cm}
and  Pavel Winternitz$^{4}$}

\vspace{0.5cm}
$^{1}${\it Centro de Ciencias F\'{\i}sicas
Universidad Nacional Aut\'onoma de M\'exico\\
Apartado Postal 48--3  62251 Cuernavaca, Morelos, M\'exico}\\[5mm]

\vspace{0.5cm}
$^{2}${\sl Laboratory of Theoretical Physics,
Joint Institute for Nuclear Research, \\
Dubna, Moscow Region 141980, Russia}

\vspace{0.5cm}
$^{3}$International Center for Advanced Studies,
Yerevan State University, \\
Alex Manougian 1,  375025, Yerevan, Armenia

\vspace{0.5cm}
$^{4}${\sl Centre de recherches math{\'e}matiques,
Universit{\'e} de Montr{\'e}al, \\
C. P. 6128, succ.
Centre Ville, Montr{\'e}al, Qu{\'e}bec, H3C 3J7, Canada}

\end{center}

\vspace{0.7cm}
\today

\vspace{1cm}

\begin{abstract}

\vspace{0.5cm}

A review is given of some recently obtained results on analytic
contractions of Lie algebras and Lie groups and their application to
special function theory. The contractions considered are from $O(3)$
to $E(2)$ and from $O(2,1)$ to $E(2)$, or  $E(1,1)$. The analytic
contractions provide relations between separable coordinate systems
on various homogeneous manifolds. They lead to asymptotic relations
between basis functions and overlap functions for the representations
of different groups.

\end{abstract}

%\maketitle
\makeatletter
\@addtoreset{equation}{section}
\renewcommand{\theequation}{\thesection.\arabic{equation}}
\makeatother

\newpage
\tableofcontents

\newpage
\section{Introduction}

Lie algebra contractions were introduced into physics by In{\"o}n{\"u}
and Wigner~\cite{IW} in 1953 as a mathematical expression of a
philosophical idea, namely the {\it correspondence principle}. This
principle tells us that whenever a new physical theory surplants an
old one, there should exist a well defined limit in which the results
of the old theory are recovered.
More specifically In{\"o}n{\"u} and Wigner established a relation
between the Lorentz group and the Galilei one in which the former goes
over into the latter as the speed of light satisfies $c \to \infty$.

The theory of Lie algebra contractions (and deformations) has acquired a
life of its own. It provides a framework in which large sets of Lie
algebras can be embedded into families depending on parameters. All
algebras in such a family have the same dimension, but they are not
mutually isomorphic \cite{KIRILLOV}.

Two types of Lie algebra contractions exist in the literature. The first
are standard In\"on\"u-Wigner contractions \cite{IW,SALETAN,GILMOR}.
They can be interpreted as singular changes of basis in a given Lie algebra
$L$. Indeed, consider a basis $\{ e_1, \dots , e_n\}$ of $L$ and a
transformation $f_i = U_{ik} (\varepsilon)e_k$, where the matrix
$U$ realizing the transformation depends on some parameters $\varepsilon$.
For $\varepsilon \to 0$ (i.e.\ some, or all of the components of
$\varepsilon$ vanishing) the matrix $U(\varepsilon)$ is singular.
In this limit the commutation relations of $L$ change
(continuously) into those of a different, nonisomorphic,
Lie algebra $L'$.

More recently, "graded contractions"  have been introduced
\cite{PAT1,PAT2,PAT3}. They are more general than the In\"on\"u-Wigner
ones and can be obtained by introducing parameters modifying the
structure constants of a Lie algebra $L_1$ in a manner respecting a
certain grading and then taking limits when these parameters go to zero.

It is well known that there exists an intimate relationship between the
theory of special functions and Lie group theory, well presented
in the books of Vilenkin \cite{VILEN1}, Talman \cite{TALMAN} and Miller
\cite{MILLER1}. In fact all properties of large classes of special functions
can be obtained from the representation theory of Lie groups, making use
of the fact that the special functions occur as basis functions of
irreducible representations, as matrix elements of transformation
matrices, as Clebsch-Gordon coefficients, or in some other guise.
Recently, the class of functions treatable by group theoretical and
algebraic methods has been extended to the so called $q$-special functions
that have been related to quantum groups \cite{EXTON,GASPER,VINET}.

One very fruitful application of Lie theory in this context is the
algebraic approach to the separation of variables in partial differential
equations \cite{WF,WLS,MPST1,MILLER21,KALNINS1,MPW}. In this approach
separable coordinate systems (for Laplace-Beltrami, Hamilton-Jacobi and
other invariant partial differential equations) are characterized by
complete sets of commuting second order operators. These lie in the
enveloping algebra of the Lie algebra of the isometry group, or in some
cases of the conformal group, of the corresponding homogeneous space.

A question that up to the last few years has received little attention
in the literature, is that of connections between the separation of
variables in different spaces, e.g. in homogeneous spaces of different
Lie groups. In particular, it is of interest to study the behavior of
separable coordinates, sets of commuting operators and the corresponding
separated eigenfunctions under deformations and contractions of the
underlying Lie algebras.

A recent series of papers
\cite{IPSW4,IPSW3,IPSW2,IP1,IP2,KMP1,IPSW1,PSWW1,IPSW5,IPSW6}
has been devoted to this new aspect of the theory of Lie algebra and
Lie group contractions: the relation between the separation of variables
in spaces of nonzero constant curvature and in flat spaces.
The curved spaces were realized as spheres $S_n \sim O(n+1)/O(n)$,
Lorentzian hyperboloids $H_n\sim O(n,1)/O(n)$, or $O(n,1)/O(n-1,1)$.
The flat spaces where either Euclidean $E_n$, or pseudo-Euclidean
$E(n-1,1)$ ones. The curved and flat spaces were related by a
contraction of their isometry groups and the corresponding isotropy
groups of the origin.

The essential point of these articles was the introduction of "analytic
contractions". The contraction parameter is $R$ which is either the
radius of the sphere $S_n$, or the corresponding quantity
$x_0^2-x_1^2 -...- x_n^2 = R^2$ for the hyperboloid. The contractions
are "analytic" because the parameter $R$ figures not only in the structure
constants of the original Lie algebra, but also in the coordinate systems,
in the operators of the Lie algebra, in the invariant operators
characterizing the coordinate systems, in the separated eigenfunctions
and the eigenvalues.

Once the parametrization displaying the parameter $R$ is established, it
is possible to follow the contraction procedure $R\to\infty$ explicitly
for all quantities. For the Lie algebra realized by vector fields, the
Laplace -Beltrami operators, the second order operators in the enveloping
algebras, characterizing separable systems of coordinates, the separated
ordinary differential equations, the eigenfunctions and the coefficients
of the interbases expansions.

For two-dimensional spaces all types of coordinates where considered.
for example, contractions of $O(3)$ to $E(2)$ relate elliptic coordinates
on $S_2$ to elliptic and parabolic coordinates on $E_2$. They also relate
spherical coordinates on $S_2$ to polar and Cartesian coordinates on $E_2$
(\cite{IPSW4,IPSW2,IP2}). Similarly, all 9 coordinate systems on the
$H_2$ hyperboloid can be contracted to at least one of the four systems
on $E_2$, or one of the 10 separable systems on $E_{1,1}$ \cite{IPSW3,IP1}.

Contractions from $S_n$ to $E_n$ where considered for subgroup type
coordinates in Ref. \cite{KMP1,IPSW1,IPSW5}, for subgroup type coordinates
and certain types of elliptic and parabolic ones.

The main application of analytic contractions in this context is to
derive special function identities, specially asymptotic formulas. Among
other possible applications we mention the theory of finite dimensional
integrable and superintegrable systems \cite{CNOR,HRW} .

    In this paper we restrict ourselves to two-dimensional spaces of
constant curvature. The organization of the article corresponds to the
table of contents.

\section{Separation of variables in two dimensional spaces of constant
curvature}

\subsection{Operator approach to the separation of variables}

Let us first consider a quite general two dimensional Riemannian,
pseudo-Riemannian or complex Riemannian spaces with a metric
%============================================================
\begin{eqnarray}
\label{METRIC1}
ds^2= g_{ik}\, du^i du^k,
\qquad
u=(u^1, u^2).
\end{eqnarray}
%============================================================
In this space we introduce a classical free Hamiltonian
%============================================================
\begin{eqnarray}
H = g_{ik} (u) p_i p_k,
\end{eqnarray}
%============================================================
where $p_i = - \partial H/\partial u_i$ are the momenta classically
conjugate to the coordinates $u^i$. We also introduce the corresponding
Laplace-Beltrami operator
%============================================================
\begin{eqnarray}
\label{LB-DELTA1}
\Delta_{LB} =  \frac{1}{\sqrt g}\frac{\partial}{\partial u^i}
{\sqrt{g}} g^{i k} \frac{\partial}{\partial u^k}.
\end{eqnarray}
%============================================================
We will be interested in two related question:

1. What are the quadratic polynomials on phase space
%============================================================
\begin{eqnarray}
\label{INT-QUAD0}
Q = a^{ik}(u) p_i p_k
\end{eqnarray}
%============================================================
that Poisson commute with the Hamiltonian
%============================================================
\begin{eqnarray}
\{Q, H\} = \sum_{i=1}^2 \left(\frac{\partial Q}{\partial u_i}
\frac{\partial H}{\partial p_i} - \frac{\partial Q}{\partial p_i}
\frac{\partial H}{\partial u_i}\right) = 0.
\end{eqnarray}
%============================================================
In other words, when do quadratic (in the momenta) integrals of motion
exist? Respectively what are the second order Hermitian operators
%============================================================
\begin{eqnarray}
\label{INT-QUAD1}
Q = \{ a^{ik}(u) \partial_{u_i} \partial_{u_k}\}
\end{eqnarray}
%============================================================
(where the bracket denotes symmetrization) that Lie commute with the
Laplace-Beltrami operator
%============================================================
\begin{eqnarray}
[Q, H] = Q H - H Q = 0 ?
\end{eqnarray}
%============================================================

2. Do the Hamilton-Jacobi and Laplace-Beltrami equations in the
considered space allow the separation of variables, and if so, how do we
classify and construct separable coordinates? By separation of
variables for the Hamilton-Jacobi equation we mean additive separation
%============================================================
\begin{eqnarray}
g_{ik} \frac{\partial S}{\partial_{u^i}}
\frac{\partial S}{\partial_{u^k}} &=& \lambda
\\
S &=& S_1(u^1, \lambda, \mu) + S_2(u^2,\lambda,\mu).
\end{eqnarray}
%============================================================
For the Laplace-Beltrami operator we have in mind multiplicative
separation
%============================================================
\begin{eqnarray}
\label{LB-HELM1}
\Delta \Psi = \lambda \Psi,
\qquad
\Psi = \Psi_1(u^1, \lambda, \mu) \Psi_2(u^2,\lambda,\mu).
\end{eqnarray}
%============================================================
In both case $\lambda$ and $\mu$ are the separation constants.

In this review article we shall mainly be interested in Laplace-Beltrami
operators in different spaces. However, some aspects of separation
are simpler to discuss for the Hamilton-Jacobi equation. In
two-dimensional Riemannian space the two equations separate in the
same coordinate systems.

The existence of integrals of motion that are either linear or
quadratic in the momenta was analyzed by Darboux \cite{DARBY} and
Koenigs \cite{KOENIGS} in a note published in Volume 4 of Darboux's
lectures. In particular it was shown that a metric (\ref{METRIC1})
can allow $0,1,2,3$ or $5$ linearly independent quadratic integrals.
The case of $5$ quadratic integrals occurs if and only if the metric
corresponds to a space of constant curvature. In this case the second
order integrals are "reducible". That means that the metric allows
precisely three first order integrals
%============================================================
\begin{eqnarray}
L_{i} =  a_i(u) p_1 + b_i(u) p_2 ,
\qquad u=(u_1, u_2),
\qquad i=1,2,3
\end{eqnarray}
%============================================================
and all second order integrals are expressed as second order polynomials
(with constant coefficients) in terms of the first order ones:
%============================================================
\begin{eqnarray}
Q = \sum_{i,k=1}^3 A_{ik} L_{i} L_k
\qquad
A_{ik}=constant
\end{eqnarray}
%============================================================
If the polynomial $Q$ is the square of a first order operator $L$,
then it will provide a subgroup type coordinate.
This is best seen by considering the corresponding first order
operator
%============================================================
\begin{eqnarray}
\label{VECTOR1}
X = \xi (u_1, u_2) \partial_{u_1} + \eta (u_1, u_2) \partial_{u_1}
\end{eqnarray}
%============================================================
that generates a one dimensional subgroup of the isometry group $G$.
>From $(u_1, u_2)$ we transform to the new coordinates $(v_1, v_2)$
"straightening out" the vector field (\ref{VECTOR1}) to the form
%============================================================
\begin{eqnarray}
\label{VECTOR2}
X = \partial_{v_1}
\end{eqnarray}
%============================================================
Then $v_1$ will be an ignorable variable. The complementary
variable $v_2= \phi (u_1, u_2)$ can be replaced by an arbitrary
function of $v_2$, the ignorable variable $v_1$ can be replaced by
$f(v_1) + g(v_2)$ where both $f$ and $g$ are arbitrary. The separable
coordinates are $v_1$ and $v_2$ (with the above mentioned
arbitrariness).

Now let us assume that an irreducible quadratic integral $Q$ as
in (\ref{INT-QUAD0}) is known for a considered metric (\ref{METRIC1})
(that is, $Q$ is not square of a linear integral).
We can then impose that two equations be satisfied simultaneously.
In the classical case they are
%============================================================
\begin{eqnarray}
\label{SEPARAB-EQ1}
g_{ik} \frac{\partial S}{\partial_{u^i}}
\frac{\partial S}{\partial_{u^k}} = \lambda,
\qquad
a_{ik} \frac{\partial S}{\partial_{u^i}}
\frac{\partial S}{\partial_{u^k}} = \mu.
\end{eqnarray}
%============================================================
Similarly, we can consider the quantum mechanics of a free particle
in such a space and write two simultaneous equation
%============================================================
\begin{eqnarray}
\label{SEPARAB-EQ2}
{\hat H} \Psi = \left(\frac{1}{\sqrt g}\frac{\partial}{\partial u^i}
{\sqrt{g}} g^{i k} \frac{\partial}{\partial u^k}\right) \Psi
= \lambda \Psi,
\qquad
{\hat Q} \Psi = \left(\frac{1}{\sqrt g}\frac{\partial}{\partial u^i}
{\sqrt{g}} a^{i k} \frac{\partial}{\partial u^k}\right) \Psi
= \mu \Psi.
\end{eqnarray}
%============================================================
Separable coordinates for the two systems (\ref{SEPARAB-EQ1})
and (\ref{SEPARAB-EQ2}) are obtained by simultaneously transforming
${\hat H}$ and ${\hat Q}$ to a standard form, in which the matrices
$g_{ik}$ and $a_{ik}$ of (\ref{SEPARAB-EQ1}) (and (\ref{SEPARAB-EQ2}))
are diagonal. This can be done by solving the characteristic
equation
%============================================================
\begin{eqnarray}
\label{CHARAC-1}
| a_{ik}- \rho \, g_{ik}| = 0.
\end{eqnarray}
%============================================================
If two distinct roots $\rho_1$ and $\rho_2$ exist, they will provide
separable coordinates, at least over the field of complex number. If we
are considering real spaces, then it may happen that $\rho_1$ and
$\rho_2$ are real only in part of the space and do not parametrize the
entire space. We will see below that this indeed happens for instance
in the pseudo-euclidean plane $E_{1,1}$.

The roots $\rho_1$ and $\rho_2$ can be replaced by any functions
$u=u(\rho_1)$, $v=v(\rho_2)$. This freedom can be used to transform $H$
and $Q$ simultaneously to the form
%============================================================
\begin{eqnarray}
\label{SEPARAT-1}
H = \frac{1}{\alpha(v)+\beta(u)} \,(p^2_u + p^2_v) = \lambda,
\qquad
Q = \frac{1}{\alpha(v)+\beta(u)} \,(\beta (u) p^2_u - \alpha(v) p^2_v)
= \mu.
\end{eqnarray}
%============================================================
The Hamiltonian $H$ in (\ref{SEPARAT-1}) is in its Liouville form
\cite{LIOUVILLE}. The separated equations are
%============================================================
\begin{eqnarray}
\label{SEPARAT-2}
\alpha H + Q = \alpha \lambda + \mu,
\qquad
\beta H - Q = \beta \lambda - \mu
\end{eqnarray}
%============================================================
for the Hamilton-Jacobi equation and similarly
%============================================================
\begin{eqnarray}
\label{SEPARAT-3}
(\alpha {\hat H} + {\hat Q})\Psi  =  (\alpha\lambda + \mu)\Psi,
\qquad
(\beta {\hat H} - {\hat Q})\Psi  =  (\beta \lambda - \mu) \Psi
\end{eqnarray}
%============================================================
for the Laplace-Beltrami equation.

Let us now restrict ourselves to two-dimensional spaces $M$ of
constant curvature, that is to the Euclidean plane $E_2$, pseudo-Euclidean
plane $E_{1,1}$, sphere $S_2$ and two, or one- sheeted hyperboloid $H_2$.
Each of these has a three-dimensional isometry group $G$. The Lie
algebra $L$ of $G$ has in each case a standard basis which we denote
$\{X_1, X_2, X_3 \}$.

The Laplace-Beltrami operator ${\hat H} = \Delta_{LB}$ (\ref{LB-DELTA1})
is in each case proportional to the Casimir operator ${\hat C}$ of the
Lie algebra $L$. The operator $Q$ commuting with $H$ will have the form
%============================================================
\begin{eqnarray}
\label{INT-MOTION2}
{\hat Q} = \sum_{i,k=1}^3 A_{ik} X_{i} X_k
\qquad
A_{ik}=A_{ki}
\end{eqnarray}
%============================================================
where $A$ is constant matrix. Let $g\in G$ be an element of the Lie
algebra of the isometry group of the considered space. Let us rewrite
eq. (\ref{INT-MOTION2}) in matrix form
%============================================================
\begin{eqnarray}
\label{INT-MOTION3}
{\hat Q} = u^{T} A u,
\qquad
u^{T} = \{X_1, X_2, X_3 \}
\end{eqnarray}
%============================================================
The transformation $g$ acting on the space $M$ induces a transformation
$u'= g u$ on the Lie algebra $L$. The Casimir operator $\Delta_{LB}$
stays invariant, but ${\hat Q}$ transforms to
%============================================================
\begin{eqnarray}
\label{INT-MOTION4}
{\hat Q} = u^{'T} g^{T} A  g u^{'}.
\end{eqnarray}
%============================================================
Thus, for spaces of constant curvature a classification of
operators ${\hat Q}$ commuting with ${\hat H}$ reduces to a
classification of symmetric matrices $A=A^{T}$ into equivalence
classes under the congruence transformation
%============================================================
\begin{eqnarray}
\label{INT-MOTION5}
A' = g^{T} A  g,
\qquad
g\in G.
\end{eqnarray}
%============================================================
This problem, as we shall see below, can be reduced to that of
classifying elements of Jordan algebras into equivalence classes.

Furthermore, the operator ${\hat C}$ of $L$ can also be written
in the form
%============================================================
\begin{eqnarray}
\label{INT-MOTION6}
{\hat C} = u^{T} C u,
\qquad
C = \,
\left(\begin{array}{ccc}
c_1&&\\
&c_2&\\
&&c_3\\
\end{array}
\right)
\end{eqnarray}
%==================================================================
Two matrices $A$ and ${\tilde A}$ will give equivalent coordinate
systems, if they satisfy
%============================================================
\begin{eqnarray}
\label{INT-MOTION7}
{\tilde A} = \lambda g^{T} A g + \mu C,
\qquad \lambda \not= 0,
\end{eqnarray}
%============================================================
where $\lambda$ and $\mu$ are real constants.

\subsection{Separable Coordinate Systems in the Euclidean Plane.}

The Lie algebra of the isometry group E(2) is given by
%============================================================
\begin{eqnarray}
L = x_2\partial_{x_1} - x_1\partial_{x_2},
\,\,\,\,
P_1 = \partial_{x_1},
\,\,\,\,
P_2 = \partial_{x_2}.
\end{eqnarray}
%============================================================
The operator ${\hat Q}$ of eq.(\ref{INT-MOTION2}) will in this case be
%============================================================
\begin{eqnarray}
{\hat Q} = a L^2 + b_1 (L P_1+P_1 L) + b_2 (L P_2+P_2 L) + c_1 P_1^2
+ c_2 P_2^2 + 2 c_3 P_1 P_2.
\end{eqnarray}
%============================================================
An $E(2)$ transformation matrix will be written as
%============================================================
\begin{eqnarray}
\label{INT-MOTION8}
g = \,
\left(\begin{array}{cc}
1&\xi^{T}\\
0&{\rm R}\\
\end{array}
\right),
\qquad
\xi^{T} = (\xi_1, \xi_2),
\quad
{\rm R}\in \bbbr^2,
\quad
{\rm R}^{T} {\rm R} = I
\end{eqnarray}
%==================================================================
The matrix  $A$ of equation (\ref{INT-MOTION3}) is
%============================================================
\begin{eqnarray}
\label{INT-MOTION9}
A = \,
\left(\begin{array}{cc}
a&b^{T}\\
b&S\\
\end{array}
\right),
\qquad
S = \,
\left(\begin{array}{cc}
c_1&c_3\\
c_3&c_2\\
\end{array}
\right),
\qquad
b^{T} = (b_1, b_2),
\end{eqnarray}
%==================================================================
and $C$ of (\ref{INT-MOTION6}) is
%============================================================
\begin{eqnarray}
\label{INT-MOTION10}
C = \,
\left(\begin{array}{ccc}
0&&\\
&1&\\
&&1\\
\end{array}
\right),
\end{eqnarray}
%==================================================================
since the Casimir operator is
%============================================================
\begin{eqnarray}
\label{CASIMIR-1}
C = \Delta = P_1^2 + P_2^2.
\end{eqnarray}
%==================================================================
The transformation (\ref{INT-MOTION7}) with $\lambda = 1$ has two
invariants in the space of symmetric matrices $A$, namely
%============================================================
\begin{eqnarray}
\label{CASIMIR-2}
I_1 = a,
\qquad
I_2 = \{[a(c_1-c_2) - (b_1^2-b_2^2)]^2 +
4 (ac_3-b_1b_2)^2\}^\frac{1}{2}
\end{eqnarray}
%==================================================================
Correspondingly, the operator ${\hat Q}$ can be transformed into one
of four canonical forms
%============================================================
\begin{eqnarray}
\label{CAR1}
1) \,\,\, I_1 = 0, \,\,\,\, I_2 = 0
\qquad
Q_C &=& P_1^2; \\
\label{POL1}
2) \,\,\, I_1 \not= 0, \,\,\,\, I_2 = 0
\qquad
Q_R &=& L^2; \\
\label{PAR1}
3) \,\,\, I_1 = 0, \,\,\,\, I_2 \not= 0
\qquad
Q_P &=& L P_2 + P_2 L;\\
\label{ELIP1}
4) \,\,\, I_1 \not= 0, \,\,\,\, I_2 \not= 0
\qquad
Q_E &=& L^2 - {D^2} P_2^2,
\qquad
D^2 =\frac{I_2}{I_1^2}
\end{eqnarray}
%============================================================

The first two forms correspond to subgroup type coordinates.
Thus, $Q_c$ of (\ref{CAR1}) corresponds to Cartesian coordinates,
in which $P_1 = \partial_x$ (and also $P_2 = \partial_y$) is already
straightened out. Both $x$
and $y$ are ignorable. The second, $Q_R$ corresponds to polar
coordinates
%=================================================================
\begin{equation}
x=r\cos\phi\enspace,
\qquad
y=r\sin\phi\enspace
\end{equation}
%=================================================================
in which $L = \partial_\phi$ is straightened out so that $\phi$ is
an ignorable variable.

The coordinates corresponding $Q_P$ of (\ref{PAR1}) are the parabolic
coordinates
%=================================================================
\begin{equation}
\label{PAR2}
x =  \frac{1}{2}(u^2 - v^2),
\qquad
y = u v.
\end{equation}
%=================================================================
Equivalently, if we take ${\tilde Q}_P = L P_1 + P_1 L$ the
prescription (\ref{CHARAC-1}) leads to
%=================================================================
\begin{equation}
\label{PAR3}
x =  u v,
\qquad
y = \frac{1}{2}(u^2 - v^2).
\end{equation}
%=================================================================
Finally, $Q_E$ of eq. (\ref{ELIP1}) corresponds to elliptic coordinates
in the plane. They can be written as
%====================================================================
\begin{eqnarray}
\label{ELLTRIG1}
x  = D \cosh\xi \cos\eta,
\qquad
y  = D\sinh\xi\sin\eta.
\end{eqnarray}
%====================================================================

%%%%%%%%%%%%%%%%%%%%%%%%%%%%%%%%%%%%%%%%%%%%%
\begin{table}
\begin{center}
{\bf Table 1:} Orthogonal coordinate systems on two-dimensional
Euclidean plane $E_2$

\vspace{0.5cm}
%\noindent
\begin{tabular}{|l|l|l|c}\hline
Coordiante systems        &Integrals of motion & Solution of     \\
                         &                    & Helmholtz equation         \\
\hline
I.~Cartesian  $x, y$          &&Exponential functions                     \\
$-\infty < x,y < \infty $     &$Q_C=P_1^2-P_2^2$                 &              \\
\hline
II.~Polar                 &           &Product of  \\
$x = r\cos\varphi$, \, $y = r\sin\varphi $         &$Q_R=L^2$&Bessel function\\
$0 \leq r < \infty, \, 0 \leq \varphi \leq 2\pi $  &&and exponential   \\
\hline
III.~Parabolic                                 &&Product of two                         \\
$x=\frac{u^2-v^2}{2}$, \, $y = uv$  &$Q_P=L P_2+ P_2 L$&parabolic cylinder \\
$0 \leq u < \infty ,-\infty < v < \infty $        &&functions  \\
\hline
IV.~Elliptic             &
&Product of periodic  \\
$x=D\cosh\xi\cos\eta $,
&$Q_{E}=L^2 - D^2 P_2^2$&and nonperiodic Mathieu\\
$y=D\sinh\xi\sin\eta $   && functions\\
$0 \leq \xi < \infty, \, 0 \leq \eta < 2\pi$   &&     \\
\hline
\end{tabular}
\end{center}
\end{table}

%%%%%%(Table 2)%%%%%%%%%%%%%%%%%%%%%%%%%%%%%%%%%%%%%%%%%%%%%%%%%%
\begin{table}
\begin{center}
{\bf Table 2:} Orthogonal coordinate systems on the two-dimensional
pseudo-Euclidean plane $E_{1,1}$

\vspace{0.5cm}
\noindent

\begin{tabular}{|l|l|l|c}\hline
Coordinate system &Integrals of motion  &Solution of \\
                  &                    &Helmholtz equation          \\
\hline
I.~Cartesian             &$Q_C = P_0 P_1$&Product of exponentials \\
 $t,\, x$                          &&\\
\hline
II.~Pseudo-polar            &   & Product of Bessel  \\
$t=r \cosh \tau_2$,\, $x=r \sinh \tau_2$ &$Q_S=K^2$&function and exponential \\
$r \geq 0$, $-\infty<\tau_2<\infty$      &&\\
\hline
III.~Parabolic of type I             &&Product of parabolic\\
$t=\frac{1}{2}(u^2+v^2)$,\, $x= uv$ &$Q_P^{I}= \{P_1, K\}$&cylinder function\\
$v \geq 0$, $-\infty<u<\infty$          && for $t^2-x^2>0$ \\
\hline
IV.~Parabolic of type II             &&Product of parabolic\\
$t = uv$, \, $x=\frac{1}{2}(u^2+v^2)$,\,  &$Q_P^{II}= \{P_0, K\}$& cylinder function\\
$v \geq 0$, $-\infty<u<\infty$          && for $x^2-t^2>0$ \\
\hline
V.~Parabolic of type III             &&Products of two linear\\
$t=\frac{1}{2}(\eta-\zeta)^2-(\eta+\zeta)$,    &$Q_P^{III}=\{P_0, K\}+\{P_1, K\}$&combinations of Airy\\
$x=\frac{1}{2} (\eta-\zeta)^   2 + (\eta+\zeta)$, &$+(P_0-P_1)^2$&functions for $x+t>0$ \\
$-\infty < \eta, \zeta < \infty$    &                            &\\
\hline
VI.~Hyperbolic of type I  &&Product of Mathieu equation\\
$t=\frac{l}{2}\left(\cosh\frac{\eta-\zeta}{2} +
\sinh\frac{\eta+\zeta}{2}\right)$       &$Q_H^I=K^2-l^2 P_0 P_1$&solutions with argument \\
$x=\frac{l}{2}\left(\cosh\frac{\eta-\zeta}{2}
-\sinh\frac{\eta+\zeta}{2}\right)$   &&displaced by $i\pi/2$\\
$-\infty<\eta,\zeta <\infty$&&\\
\hline
VII.~Hyperbolic of type II      &&Product of two solutions\\
$t= \left(\sinh(\eta-\zeta) + e^{\eta+\zeta}\right)$
&$Q_H^{II}=K^2+ (P_1+P_2)^2$&of Bessel's equation,\\
$x= \left(\sinh(\eta-\zeta) - e^{\eta+\zeta}\right)$  &&one with real and one\\
$-\infty<\eta,\zeta <\infty$                           &&with imaginary arguments\\
\hline
VIII.~Hyperbolic 0f type III  &&Product of two solutions\\
$t= \left(\cosh(\eta-\zeta) + e^{\eta+\zeta}\right)$
&$Q_H^{III}=K^2-(P_1+P_2)^2$&of Bessel's equation\\
$x= \left(\cosh(\eta-\zeta) - e^{\eta+\zeta}\right)$ &&\\
$-\infty<\eta,\zeta <\infty$   &&\\
\hline
\end{tabular}
\end{center}
\end{table}

%\newpage
\begin{table}
\begin{center}
{\bf Table 2:} (Continue)

\noindent
\vspace{0.5cm}

\begin{tabular}{|l|l|l|c}\hline
Coordinate system &Integrals of motion  &Solution of \\
                  &                    &Helmholtz equation          \\
\hline
IX.~Elliptic of type I        &&Product of two solutions\\
$t= D\sinh \eta \cosh \zeta$, \, $x= D\cosh \eta \sinh \zeta$
&$Q_E^{I}=K^2+D^2 P_1^2$&of the nonperiodic Mathieu\\
$-\infty < \eta, \zeta < \infty$     &&equation\\
\hline
X.~Elliptic  of type II &&Product of two solutions\\
(i) $t= d\cosh \eta \cosh \zeta$, \, $x= d\sinh \eta \sinh \zeta $
&$Q_E^{II}=K^2-d^2 P_1^2$&(i) of the nonperiodic  \\
$-\infty<\eta <\infty$, $\zeta \geq 0$&&Mathieu equation\\
(ii) $t= d\cos
\eta \cos \zeta$, \, $x= d\sin \eta \sin \zeta$ &&(ii) of the periodic\\
$0<\eta<2\pi, 0\leq \zeta < \pi$                &&Mathieu equation\\
&&\\
\hline
\end{tabular}
\end{center}
\end{table}

%--------------------------------------------------------------------

\subsection{Separable Coordinate Systems in the Pseudo-Euclidean Plane}

The Lie algebra of the isometry group $E(1,1)$ can be represented by
%============================================================
\begin{eqnarray}
\label{MINKOW-1}
K   = \left(t\partial_{x} + x\partial_{t}\right),
\qquad
P_0 = \partial_{t},
\qquad
P_1 = \partial_{x}.
\end{eqnarray}
%============================================================
The second order operator (\ref{INT-MOTION2}) is
%============================================================
\begin{eqnarray}
\label{MINKOW-2}
{\hat Q} = a K^2 + b_0 (K P_0+ P_0 K) +
b_1 (K P_1 + P_1 K) + c_0 P_0^2 + c_1 P_1^2 + 2c_2 P_0 P_1.
\end{eqnarray}
%============================================================
Equivalently, the matrix $A$ of (\ref{INT-MOTION5}) is
%============================================================
\begin{eqnarray}
\label{MINKOW-3}
A = \,
\left(\begin{array}{cc}
a&b^{T}\\
b&c\\
\end{array}
\right),
\qquad
C = \,
\left(\begin{array}{cc}
c_0&c_2\\
c_2&c_1\\
\end{array}
\right),
\qquad
b^{T} = (b_0, b_1)
\end{eqnarray}
%==================================================================
We will classify the operators ${\hat Q}$ into conjugate classes and
the action of the group $E(1,1)$, including the reflactions
%============================================================
\begin{eqnarray}
\label{MINKOW-4}
\Pi_0: \,\,\,\, (x,t) \to (x, -t),
\qquad
\Pi_1: \,\,\,\, (x,t) \to (-x, t).
\end{eqnarray}
%==================================================================
An element of $E(1,1)$, acting on the Lie algebra $(K,P_0,P_1)$ can
be represented as
%============================================================
\begin{eqnarray}
\label{MINKOW-5}
g = \,
\left(\begin{array}{cc}
1&\xi^{T}\\
0&\Lambda\\
\end{array}
\right),
\qquad
\xi^{T} = (\xi_0, \xi_1),
\quad
\Lambda \in \bbbr^2,
\,\,\,
\Lambda^{T} J \Lambda = J
\end{eqnarray}
%==================================================================
with
%============================================================
\begin{eqnarray}
\label{MINKOW-6}
J = \,
\left(\begin{array}{cc}
1&0\\
0&-1\\
\end{array}
\right).
\end{eqnarray}
%==================================================================
The matrix A of (\ref{MINKOW-3}) is subject to the transformation
(\ref{INT-MOTION7}) and in this case we have
%============================================================
\begin{eqnarray}
\label{MINKOW-7}
A' = g^{T} A g = \,
\left(\begin{array}{cc}
a&a\xi^{T} +\beta^{T} \Lambda\\
\xi a + \Lambda^{T} \beta&
\Lambda^{T} C \Lambda +  \Lambda^{T} \beta \xi^{T} +
\xi \beta^{T} \Lambda + a \xi \xi^{T}\\
\end{array}
\right).
\end{eqnarray}
%==================================================================
One of the invariant of this transformation is the constant a which
can be closen to be $a=1$, or is already $a=0$.

Let us first consider $a\not=0$. Choosing $\xi=-\beta^{T} \Lambda$
and putting $a=1$, we obtain
%============================================================
\begin{eqnarray}
\label{MINKOW-8}
A' = \,
\left(\begin{array}{cc}
1&0\\
0&C'\\
\end{array}
\right),
\qquad
C' = J \Lambda^{-1} J (C-\beta \beta^{T}) \Lambda.
\end{eqnarray}
%==================================================================
Notice that $C'$ and $C$ are symmetric matrices, but we have
%============================================================
\begin{eqnarray}
\label{MINKOW-9}
X \equiv J (C-\beta \beta^{T}),
\qquad
J X^{T} = X J,
\end{eqnarray}
%==================================================================
that is, $X$ is an element of the Jordan algebra $jo(1,1)$. Since
$\Lambda$ is an element of the Lie group $O(1,1)$, we are faced with
a well known problem: the classification of elements of a Jordan algebra
with respect to conjugation under the corresponding Lie group.
The results are known for all classical Lie and Jordan algebras
\cite{JORDAN}, and for $jo(1,1)$ they are quite simple.
The matrix $X$ can be transformed into one of the following
%============================================================
\begin{eqnarray}
\label{MINKOW-10}
X_1 = \,
\left(\begin{array}{cc}
p&0\\
0&q\\
\end{array}
\right),
\quad
X_2 = \,
\left(\begin{array}{cc}
p&q\\
-q&p\\
\end{array}
\right),
\quad
X_3 = \,
\left(\begin{array}{cc}
p+\epsilon&\epsilon\\
-\epsilon&p-\epsilon\\
\end{array}
\right),
\quad
q>0,
\quad
\epsilon = \pm 1,
\end{eqnarray}
%==================================================================
with $p\in \bbbr$, $q\in \bbbr$.

For $a=0$, $(b_0,b_1)\not=(0,0)$ transformation (\ref{INT-MOTION7})
leads to eq. (\ref{MINKOW-7}) in which we set $a=0$. We choose
the matrix $\Lambda$ contained in $O(1,1)$ to transform $\Lambda^{T} \beta$ to
standard form and then choose $\xi$ to simplify the matrix $C$.
For $|b_0|>|b_1|$, $|b_0|<|b_1|$ and $|b_0|=|b_1|$ we can transform
$A$ into
%==================================================================
\begin{eqnarray}
\label{MINKOW-11}
A_1 &=& \,
\left(\begin{array}{ccc}
0&\epsilon\sqrt{b_0^2-b_1^2}&0\\
\epsilon\sqrt{b_0^2-b_1^2}&0&0\\
0&0&0\\
\end{array}
\right), \qquad \epsilon = \pm1
\\[2mm]
\label{MINKOW-12}
A_2 &=& \,
\left(\begin{array}{ccc}
0&0&\epsilon\sqrt{b_1^2-b_0^2}\\
0&0&0\\
\epsilon\sqrt{b_1^2-b_0^2}&0&0\\
\end{array}
\right), \qquad \epsilon = \pm1
\\[2mm]
\label{MINKOW-13}
A_3 &=& \,
\left(\begin{array}{ccc}
0&1&1\\
1&\gamma&-\gamma\\
1&-\gamma&\gamma\\
\end{array}
\right),
\qquad
\gamma =0,1
\end{eqnarray}
%==================================================================
respectively.

Finally, for $a=b_0=b_1=0$, $C\not=0$ we can use $\Lambda$ to transform
$C$ into one of its standard forms $J X_i$, $i=1,2,3$ with $X_i$ as
in (\ref{MINKOW-10}).

Thus, we have obtained a classification of matrices $A$ that determine
the operator ${\hat Q}$. Let us now list the corresponding operators.
We first notice that if $a=b_1=b_2=0$. The corresponding operator
${\hat Q}$ is in the enveloping algebra of a maximal Abelian subalgebra
of $e(1,1)$, namely $(P_0, P_1)$. Similarly, for $a=1$ and $X=X_1$
in eq. (\ref{MINKOW-10}) with $p=q$ we find that $Q=K^2$ is in the
enveloping algebra of a different maximal abelian subalgebra of $e(1,1)$,
namely $o(1,1)$ (generated by $K$). These two cases correspond to subgroup
type coordinates. The other ones to nonsubgroup type.

The list of operators must be further simplifield by linear combinations
with the Casimir operator
%============================================================
\begin{eqnarray}
\label{CASIMIR-21}
C = P_0^2 - P_1^2.
\end{eqnarray}
%==================================================================
Finally, we obtain a representative list of 11 second order operators
in the enveloping algebra of the Lie algebra $e(1,1)$.
%============================================================
\begin{eqnarray}
\label{CAR11}
Q_1 (a,b) &=& a(P_0^2+P_1^2) + 2b P_0 P_1;
\nonumber\\
&& (a,b) = (1,0), (1,1), {\mbox {or}} (0,1)
\nonumber\\
Q_2 &=& K^2;
\nonumber\\
Q_3 &=& K P_1 + P_1 K;
\nonumber\\
Q_4 &=& K P_0 + P_0 K;
\nonumber\\
Q_5 &=& K (P_0 + P_1) + (P_0 + P_1) K;
\nonumber\\
Q_6 &=& K (P_0 + P_1) + (P_0 + P_1) K + (P_0-P_1)^2;
\\
Q_7 &=& K^2 - {l^2} P_0 P_1;  \qquad l > 0
\nonumber\\
Q_8 &=& K^2 - {D^2} P_1^2;    \qquad D > 0
\nonumber\\
Q_9 &=& K^2 - {d^2} P_1^2;    \qquad d > 0
\nonumber\\
Q_{10} &=& K^2 + (P_0+P_1)^2;
\nonumber\\
Q_{11} &=& K^2 - (P_0+P_1)^2.
\nonumber
\end{eqnarray}
%============================================================

To obtain separable coordinates we proceed as in Section 2.1.

1. The operator $Q_1 (a,b)$ for any $a$ and $b$, corresponds to
Cartesian coordinates ($t,x$), since the operators that are really
diagonalized are $P_0$ and $P_1$ (they correspond to a maximal
Abelian subalgebra $\{P_0, P_1\}\in e(1,1)$.

2. The operator $Q_2 = K^2$ also corresponds to subgroup type
coordinates, namely pseudopolar coordinates
%=================================================================
\begin{equation}
\label{PSEUDO-SPHER1}
t=r\cosh\alpha,
\qquad
x=r\sinh\alpha
\end{equation}
$$
0\leq r < \infty, \qquad \infty < \alpha < \infty.
$$
%=================================================================
These coordinates only cover part of the pseudo-Euclidean plane, since
we have $t^2-x^2 = r^2 \geq 0$. By interchanging $t$ and $x$ in
(\ref{PSEUDO-SPHER1}) we can parametrize the part with
$t^2-x^2 = - r^2$

The operators $Q_3, ..... Q_{11}$ can lead to separable coordinates via
the algorithm of eq.~(\ref{CHARAC-1}). Two problems can and do occur.
The first is that the roots of eq.~(\ref{CHARAC-1}) may coincide:
$\rho_1 = \rho_2$. Then we do not obtain separable coordiantes.
This happens in precisely one case, namely that of the operator
$Q_5$.

To other problem that may occur is that the eigenvalues $\rho(t,x)$
may be complex in at least part of the ($x,t)$ plane. This part of the
plane will then not be covered by the corresponding coordinates
($\rho_1,\rho_2$).

The results of this analysis are presented in {\it Table 2} and
essentially agree with those of Kalnins \cite{KALNINS2}.

\subsection{The systems of coordinates on $S_2$}

The Lie algebra of isometry group $O(3)$ is given by
%============================================================
\begin{eqnarray}
L_i = - \epsilon_{ikj}u_{k}\frac{\partial}{\partial u_j}
,\,\,\,\,\,\,
[L_i , L_k] =  \epsilon_{ikj} L_j,
\,\,\,\,\,\,
i,k,j = 1,2,3.
\end{eqnarray}
%============================================================
where $u_i$ are the Cartesian coordinates in the ambient Euclidean
space $E_3$. On the sphere $S_2$ we have $u_1^2 + u_2^2 + u_3^2 = R^2$.
The Casimir operator is
%============================================================
\begin{eqnarray}
C = R^2 \Delta_{LB} = L_1^2 + L_2^2 +L_3^2
\end{eqnarray}
%============================================================
and the Laplace-Beltrami equation (\ref{LB-HELM1}) for $S_2$ has
the form
%===============================================================
\begin{eqnarray}
\label{LB-HELM2}
\Delta_{LB} \Psi = - \frac{\ell(\ell+1)}{R^2} \Psi,
\,\,\,\,\,
\Psi_{lk}(\alpha, \beta) = \Xi_{lk}(\alpha) \Phi_{lk}(\beta),
\end{eqnarray}
%===============================================================
where $\ell = 0,1,2,...$. The second order operator ${\hat Q}$ of
eq.~(\ref{INT-MOTION2}) is given by
%===============================================================
\begin{eqnarray}
{\hat Q} = A_{ik} L_i L_k, \,\,\,\,\,\,\,\,\, A_{ik} = A_{ki}.
\end{eqnarray}
%============================================================
The transformation matrix for $O(3)$ can be represented as
%================================================================
\begin{eqnarray}
\label{ROTAT1}
g= \left(\begin{array}{ccc}
\cos\alpha\cos\beta\cos\gamma-\sin\alpha\sin\gamma&
-\cos\alpha\cos\beta\sin\gamma-\sin\alpha\cos\gamma&
\cos\alpha\sin\beta\\
\sin\alpha\cos\beta\cos\gamma+\cos\alpha\sin\gamma&
-\sin\alpha\cos\beta\sin\gamma+\cos\alpha\cos\gamma&
\sin\alpha\sin\beta\\
-\sin\beta\cos\gamma&\sin\beta\sin\gamma&\cos\beta\\
\end{array}
\right)
\end{eqnarray}
%==================================================================
where ($\alpha,\beta,\gamma$) are the Euler angles.

The matrix $A_{ik}$  can be diagonalized to give
%============================================================
\begin{eqnarray}
\label{ENALGEB1}
{\hat Q}(a_1, a_2, a_3) \equiv Q = a_1 L_1^2 + a_2 L_2^2 + a_3 L_3^2.
\end{eqnarray}
%============================================================
For $a_1=a_2=a_3$ we have $Q \sim 0$. If two eigenvalues of
$A_{ik}$ are equal, e.g. $a_1=a_2\not=a_3$, or $a_1\not=a_2=a_3$,
or $a_1=a_3\not=a_2$ we can transform the operator $Q$ into the
operators: $Q(0,0,1) = L_3^2$, $Q(1,0,0) = L_1^2$ or $Q(0,1,0)=L_2^2$
respectively. The corresponding separable coordinates on $S_2$ are
the three types of spherical ones
%================================================================
\begin{eqnarray}
\label{COOR1}
\begin{array}{llll}
u_1 &=\, R\sin\theta\cos\varphi
    &=\, R\cos\theta'
    &=\, R\sin\theta''\sin\varphi'', \\
u_2 &=\, R\sin\theta\sin\varphi
    &=\, R\sin\theta'\cos\varphi'
    &=\, R\cos\theta'',              \\
u_3 &=\, R\cos\theta
    &=\, R\sin\theta'\sin\varphi'
    &=\, R\sin\theta''\cos\varphi''
\end{array}
\end{eqnarray}
%==================================================================
where $\varphi \in [0, 2\pi),$ $\theta \in [0, \pi]$. They correspond
to the group reduction $O(3)\supset O(2)$ and $X=L_i^2$ is invariant
under $O(2)$ and under reflections in all coordinate planes.

The $O(3)$ unitary irreducible representation matrix elements of
(\ref{ROTAT1}) results in the well-known transformation formula for
spherical functions $Y_{lm}(\theta,\varphi)$ \cite{VILEN1,VAR}
%================================================================
\begin{eqnarray}
\label{ROTAT11}
Y_{l,m'}(\theta',\varphi')
=
\sum_{m=-l}^{l}\, D_{mm'}^l
(\alpha, \beta \gamma) \, Y_{l,m}(\theta,\varphi),
\end{eqnarray}
%==========================================================
where $D_{m_1, m_2}^l (\alpha,\beta,\gamma)$ - are the Wigner
$D$-functions
%==========================================================
\bea
D_{m,m'}^\ell (\alpha, \beta, \gamma)
&=& e^{-im\alpha}\,d_{m,m'}^\ell(\beta)\,e^{-im'\gamma},
\label{BIG-DFUN}\\
d_{m,m'}^\ell(\beta)
&=& {(-1)^{m-m'}\over(m-m')!}
\sqrt{{(\ell+m)!\,(\ell-m')!\over(\ell-m)!\,(\ell+m')!}}
(\cos\onehalf\beta)^{2\ell-m+m'}
(\sin\onehalf\beta)^{m-m'}\hskip-1cm{}
\nonumber\\ & &\quad{}\times
F\left[{m-\ell,\ -m'-\ell\atop m-m'+1};
{-\tan^2}\,\onehalf\beta\right]
\label{LITTLE-DFUN}
\eea
%==========================================================
and the spherical angles ($\theta,\varphi$) and ($\theta',\varphi'$)
are  related by
%==========================================================
\begin{eqnarray}
\label{ROTAT12}
\cos\theta'
&=& \cos\theta \cos\beta + \sin\theta\sin\beta\cos(\varphi-\alpha)
\nonumber\\[2mm]
\cot(\varphi'+ \gamma)
&=& \cot(\varphi-\alpha)\cos\beta
-\frac{\cot\theta\sin\beta}{\sin(\varphi-\alpha)}
\end{eqnarray}
%================================================================
In particular, $Y_{lm}(\theta,\varphi)$ corresponding to the solution
of Laplace-Beltra\-mi equation in the systems of coordinates (\ref{COOR1})
are related by the formulas
%================================================================
\begin{eqnarray}
\label{EXP1}
Y_{l,m'}(\theta',\varphi')&=&\sum_{m=-l}^{l}D_{mm'}^l
(0,\frac{\pi}{2},\frac{\pi}{2})
Y_{l,m}(\theta,\varphi),
\\[2mm]
\label{EXP2}
Y_{l,m''}(\theta'',\varphi'')&=&\sum_{m=-l}^{l}D_{mm''}^l
(\frac{\pi}{2},\frac{\pi}{2},0)
Y_{l,m}(\theta,\varphi),
\\[2mm]
\label{EXP3}
Y_{l,m''}(\theta'',\varphi'')&=&\sum_{m'=-l}^{l}D_{m'm''}^l
(0,\frac{\pi}{2},\frac{\pi}{2})
Y_{l,m'}(\theta',\varphi').
\end{eqnarray}
%==========================================================

When all three eigenvalues $a_{i}$ are different, then the separable
coordinates in eq.~(\ref{LB-HELM2}) are elliptic ones
\cite{PAWI,LS1,LUKA1}.
These can be written in algebraic form, as
%==================================================================
\begin{eqnarray}
\label{ALGEB1}
 u_1^{2} = R^2{{(\rho_1-a_1)
 (\rho_2-a_1)}\over{(a_2-a_1)(a_3-a_1)}},\,\,\,\,
 u_2^{2} = R^2{{(\rho_1-a_2)(\rho_2-a_2)}
 \over{(a_3-a_2)(a_1-a_2)}},\,\,\,\,
 u_3^{2} = R^2{{(\rho_1-a_3)(\rho_2-a_3)}
 \over{(a_1-a_3)(a_2-a_3)}}
\end{eqnarray}
%==============================================================
with $a_1\leq \rho_1 \leq a_2 \leq \rho_2 \leq a_3$.

In trigonometric form we put
%====================================================================
\begin{eqnarray}
\rho_1 = a_1 + (a_2 - a_1)\cos^2\phi,\,\,\,\,\,\,
\rho_2 = a_3 - (a_3 - a_2)\cos^2\theta,
\end{eqnarray}
%====================================================================
and obtain
%====================================================================
\begin{eqnarray}
\label{TRIG1}
u_1 = R \sqrt{1 - k'^2\cos^2\theta} \cos\phi,\,\,\,\,\,
u_2 = R \sin\theta\sin\phi,\,\,\,\,\,
u_3 = R \sqrt{1-k^2\cos^2\phi} \cos\theta.
\end{eqnarray}
$$
0 \leq \phi < 2\pi, \,\,\,\,
0 \leq \theta \leq \pi,
$$
%=================================================================
where
%==================================================================
\begin{eqnarray}
\label{MODUL1}
k^2 = \frac{a_2 - a_1}{a_3 - a_1}
= \sin^2 f,
\,\,\,\,\,\,\,\,\,
k^{'2} = \frac{a_3 - a_2}{a_3 - a_1}
= \cos^2 f.
\,\,\,\,\,\,\,\,\,
k^2 + k^{'2} = 1.
\end{eqnarray}
%=================================================================
The Jacobi elliptic version of elliptic coordinates is obtained by
putting
%====================================================================
\begin{eqnarray}
\label{EL-SPHER11}
\rho_1 = a_1 + (a_2 - a_1)\sn^2 (\alpha, k),\,\,\,\,\,\,
\rho_2 = a_2 + (a_3 - a_2)\cn^2 (\beta, k'),
\end{eqnarray}
%====================================================================
We obtain
%====================================================================
\begin{eqnarray}
\label{EL1}
u_1 = R\, \sn(\alpha, k) \dn(\beta, k'),\,\,\,\,\,
u_2 = R\, \cn(\alpha, k) \cn(\beta, k'),\,\,\,\,\,
u_3 = R\, \dn(\alpha, k) \sn(\beta, k'),
\end{eqnarray}
$$
-K   \leq \alpha \leq K, \,\,\,\,
-2K' \leq \beta \leq 2K',
$$
%=================================================================
where $\sn(\alpha, k)$, $\cn(\alpha, k')$ and $\dn(\beta,k)$
are the Jacobi elliptic functions with modulus $k$ and $k'$,
and $K$ and $K'$ are the complete elliptic integrals \cite{BE3}.

The interfocal distance for the ellipses on the upper hemisphere
is equal to $2fR$.

%%%%%%%%%%%%%%%%%%%%%%%%%%%%%%%%%%%%%%%%%%%%%
\begin{table}
\begin{center}
{\bf Table 3:} Orthogonal coordinate systems on two-dimensional sphere
$S_2$

%%%%%%%%%%%%%%%%%%%%%%%%%%%%%%%%%%%%%%%%%%%%%%%%%%%%%%%%%%%%%%%%%%%%%%%
\vspace{0.5cm}
%\noindent
\begin{tabular}{|l|l|l|l|c}\hline
Coordinate systems             & Integrals&Solution of         &Limiting \\
                               & of motion&Helmholtz equation  &systems on $E_2$\\
\hline
I.~Spherical                 &            &Product of associated
& Polar           \\
$u_1=R\sin\theta\cos\varphi$ &$Q_S=L_3^2$ &Legendre polynomials& Cartesian\\
$u_2=R\sin\theta\sin\varphi$ &            &and exponential& \\
$u_3=R\cos\theta$            &            && \\
$\varphi \in [0, 2\pi), \theta\in [0,\pi]$ &&&               \\
\hline
II.~Elliptic                         &&Product of two&Elliptic                         \\
$u_1=R\, \sn(\alpha, k) \dn(\beta, k')$
&$Q_{E}= k'^2 L_3^2-k^2L_1^2$        &Lam\'e polynomials&Polar\\
$u_2=R\, \cn(\alpha, k) \cn(\beta, k')$&        &&Cartesian\\
$u_3=R\, \dn(\alpha, k) \sn(\beta, k')$&        &&Parabolic$^{*}$\\
$\alpha \in [-K, K]$,           &        &&\\
$\beta \in [-2K', 2K']$         &        &&\\
\hline
\end{tabular}
\end{center}

\vspace{0.1cm}
\footnotesize{$^{*}$} After rotation.
\end{table}
%%%%%%%%%%%%%%%%%%%%%%%%%%%%%%%%%%%%%%%%%%%%%%%%%%%%%%%%%%%%%%%%%%%%%%%%%%

\subsection{Systems of coordinates on $H_2$}

The isometry group for the hyperboloid $H_2$: $u_0^2-u_1^2-u_2^2 = R^2$,
where $u_i \, (i=0,1,2)$ are the Cartesian coordinates in the ambient
space $E_{2,1}$ is O(2,1). We choose a standard basis ${K_1,K_2,L_3}$
for the Lie algebra o(2,1):
%=================================================================
$$
  K_1 = -\left(u_0 {\partial_{u_2}} + u_2
  {\partial_{u_0}}\right)\!,      \
  K_2 = -\left(u_0{\partial_{u_1}} +
  u_1{\partial_{u_0}}\right)\!, \
  L_3 = - \left(u_1{\partial_{u_2}} -
  u_2{\partial_{u_1}}\right)
$$
%====================================================================
with commutation relations
%====================================================================
\begin{eqnarray}
\label{NONCOMPACT1}
\left[K_1,K_2\right] = - L_3,\,\,\,\,\,
\left[L_3,K_1\right] =   K_2,\,\,\,\,\,
\left[K_2,L_3\right] =   K_1.
\end{eqnarray}
%===================================================================
The Casimir operator is
%==================================================================
\begin{equation}
C = R^2 \, \Delta_{LB} =  K_1^2 + K_2^2 - L_3^2.
\end{equation}
%===================================================================
and the Laplace-Beltrami equation (\ref{LB-HELM1}) given by
%===============================================================
\begin{eqnarray}
\label{HE1}
\Delta_{LB} \Psi = \frac{l(l+1)}{R^2} \Psi,
\,\,\,\,\,
\Psi_{l\lambda }(\zeta_1,\zeta_2) =
\Xi_{l\lambda }(\zeta_1) \Phi_{l\lambda }(\zeta_2),
\end{eqnarray}
%===============================================================
where $\ell$ for principal series of the unitary irreducible
representations has the form
%==============================================================
\begin{eqnarray}
\ell = - \frac{1}{2} + i\rho,\,\,\,\,\, 0 < \rho < \infty
\end{eqnarray}
%===============================================================
The second order operator ${\hat Q}$ of eq.~(\ref{INT-MOTION2}):
%===============================================================
\begin{eqnarray}
{Q}= a K_1^2 + b (K_1K_2 + K_2K_1) + c K_2^2
+ d (K_1 L_3 + L_3 K_1) + e (K_2 L_3 + L_3 K_2) + f L_3^2,
\end{eqnarray}
%============================================================
can be used to classify all coordinate systems on $H_2$.  The
classification of the operators $Q$ can be reduced to a classification
of the normal forms of the elements of the Jordan algebra jo(2,1)
\cite{JORDAN}. There are 9 inequivalent forms, in one to one
correspondance with the 9 existing separable coordinate systems
\cite{WLS,OLEV,GROP4}. All the coordinate systems are orthogonal ones.

The normal forms of the operator $Q$ and the corresponding coordinates
are given in {\it Table 4}. Cases I, II, and III correspond to subgroup
type coordinates. The corresponding subgroups are $O(2)$, $O(1,1)$ and
$E(1)$, respectively. The $O(1,1)$ subgroup in the equidistant coordinates
acts in the $01$ plane. We could also have chosen the $02$ plane
(i.e. permuted $u_1$ and $u_2$).

The elliptic and hyperbolic coordinates of cases IV and V are given
in algebraic form. Equivalently, they can be expressed e.g. in terms of
Jacobi elliptic functions. This makes it possible to express the
coordinates in the ambient space directly, rather than their squares.
Indeed, if we put
%==============================================================
\begin{equation}
\vrho_1=a_1-(a_1-a_3)\dn^2(\alpha,k)\enspace,\qquad
\vrho_2=a_1-(a_1-a_2)\sn^2(\beta,k')\enspace,
\end{equation}
%==============================================================
and
%==============================================================
\begin{eqnarray}
\label{MODUL11}
k^2={a_2-a_3\over a_1-a_3}\enspace,
\qquad
{k'}^2={a_1-a_2\over a_1-a_3}\enspace,
\qquad
k^2+{k'}^2=1,
\end{eqnarray}
%==============================================================
into the expressions
%==================================================================
\begin{eqnarray}
\label{ELALGEB1}
u_0^{2} = R^2{{(\vrho_1-a_3)
(\vrho_2-a_3)}\over{(a_1-a_3)(a_2-a_3)}},\,\,\,\,
u_1^{2} = R^2{{(\vrho_1-a_2)(\vrho_2-a_2)}
\over{(a_2-a_3)(a_1-a_2)}},\,\,\,\,
u_2^{2} = R^2{{(\vrho_1-a_1)(a_1-\vrho_2)}
\over{(a_1-a_2)(a_1-a_3)}},
\end{eqnarray}
%=============================================================
we obtain the elliptic coordinates in Jacobi form
%==============================================================
\begin{equation}
u_0= R\,\sn(\alpha,k)\dn(\beta,k'), \,\,\,\,
u_1= i\, R\cn(\alpha,k)\cn(\beta,k'),  \,\,\,\,
u_2= i\, R\dn(\alpha,k)\sn(\beta,k'),
\end{equation}
$$
\alpha\in (iK',iK"+2K),  \qquad\beta\in [0, 4K').
$$
%==============================================================

%%%%%%%%%%%%%%%%%%%%%%%%%%%%%%%%%%%%%%%%%%%%%
\vspace{0.3cm}
%\begin{table}[t]
\begin{table}
\begin{center}
{\bf Table 4:} Orthogonal systems of coordinate on two-dimensional
hyperboloid  $H_{2}$

\vspace{0.5cm}

\noindent
\begin{tabular}{|l|l|l|l|c}\hline
Coordinate systems &Coordinate   &Limiting&Limiting \\
and integrals of motion &             &systems on $E_2$&systems on $E_{1,1}$      \\
\hline
I.~Pseudo-spherical       &$u_0=R\cosh\tau$&          Polar& Cartesian\\
$\tau>0,\rphi$              &$u_1=R\sinh\tau\cos\vphi$&   &   \\
$Q_S=L_3^2$                 &$u_2=R\sinh\tau\sin\vphi$&               & \\
\hline
II.~Equidistant          &$u_0=R\cosh\tau_1\cosh\tau_2$& Cartesian& Polar\\
$\tau_{1,2}\in\bbbr$        &$u_1=R\cosh\tau_1\sinh\tau_2$&           & \\
$Q_{EQ}=K_1^2$              &$u_2=R\sinh\tau_1$           &           & \\
\hline
III.~Horocyclic            &$u_0=R(\tilde x^2+ \tilde y^2+1)/2\tilde y$&Cartesian&Rectangular \\
$\tilde y>0,\tilde x\in\bbbr$  &$u_1=R(\tilde x^2+\tilde y^2-1)/2\tilde y$& &Cartesian coordi-  \\
$Q_{HO}=(K_1+L_3)^2$           &$u_2=R\tilde x/\tilde y$&        &nates rotated by $\frac{\pi}{4}$\\
&&&(nonorthogonal)\\
\hline
IV.~Elliptic     &$u_0^2=R^2\frac{(\rho_1-a_3)(\rho_2-a_3)}{(a_1-a_3)(a_2-a_3)}$&Elliptic&Elliptic I,II,III \\
$a_3<a_2<\rho_2<a_1<\rho_1$  &$u_1^2=R^2\frac{(\rho_1-a_2)(\rho_2-a_2)}{(a_1-a_2)(a_2-a_3)}$&Parabolic&Cartesian\\
$Q_{E}=L_3^2+\sinh^2f K_2^2$ &$u_2^2=R^2\frac{(\rho_1-a_1)(a_1-\rho_2)}{(a_1-a_2)(a_1-a_3)}$&Cartesian&\\
\hline
V.~Hyperbolic         &$u_0^2=R^2\frac{(\rho_1-a_2)(a_2-\rho_2)}{(a_1-a_2)(a_2-a_3)}$&Cartesian&Elliptic II       \\
$\rho_2<a_3<a_2<a_1<\rho_1$       &$u_1^2=R^2\frac{(\rho_1-a_3)(a_3-\rho_2)}{(a_1-a_3)(a_2-a_3)}$&       &Parabolic I \\
$Q_{H}=K_2^2-\sin^2\alpha L_3^2$  &$u_2^2=R^2\frac{(\rho_1-a_1)(a_1-\rho_2)}{(a_1-a_2)(a_1-a_3)}$&       &\\
\hline
VI.~Semi-hyperbolic    &$\frac{u_0^2+u_1^2}{R^2}= (1+\mu_1^2)(1+\mu_2^2)$&Parabolic&Cartesian\\
$\mu_{1,2}>0\spaceTT$       &$\frac{u_0^2-u_1^2}{R^2}=(1+\mu_1\mu_2)$      &Cartesian &\\
$Q_{SH}=-\{K_1,L_3\}$         &$u_2=R\sqrt{\mu_1\mu_2}$        & &\\
\hline
VII.~Elliptic-parabolic            &$u_0=R\dfrac{\cosh^2a+\cos^2\vtheta}{2\cosh a\cos\vtheta}$&Parabolic&Hyperbolic II\\
$a\in\bbbr,\vtheta\in(-\pi/2,\pi/2)$   &$u_1=R\dfrac{\sinh^2a-\sin^2\vtheta}{2\cosh a\cos\vtheta}$&           &\\
$Q_{EP}=(K_1+L_3)^2+K_2^2$               &$u_2=R\tan\vtheta\tanh a$&         &\\
\hline
VIII.~Hyperbolic-parabolic&$u_0=R\dfrac{\cosh^2b+\cos^2\vtheta}{2\sinh b\sin\vtheta}$&Cartesian&Hyperbolic III \\
$b>0,\vtheta\in(0,\pi)$   &$u_1=R\dfrac{\sinh^2b-\sin^2\vtheta}{2\sinh b\sin\vtheta}$&   &\\
$Q_{HP}=(K_1+L_3)^2-K_2^2$  &$u_2=R\cot\vtheta\coth b$&           &\\
\hline
IX.~Semicircular-parabolic    &$u_0=R\dfrac{(\SCPZ)^2+4}{8\xi\eta}$&Cartesian  &does not correspond\\
$\xi,\eta>0\spaceTT$          &$u_1=R\dfrac{(\SCPZ)^2-4}{8\xi\eta}$&      &to a separable  \\
$Q_{SCP}=\{K_1,K_2\}$        &$u_2=R\dfrac{\eta^2-\xi^2}{2\xi\eta}$&       &coordinate system  \\
$+ \{K_2,L_3\}$ &&&\\
\hline
\end{tabular}
\end{center}
\end{table}

\section{Contractions of the Lie algebra and Casimir operator}

\subsection{Contractions from $o(3)$ to $e(2)$}

We shall use $R^{-1}$ as the contraction parameter. To realize the
contraction explicitly, let us introduce homogeneous or Beltrami
coordinates on the sphere, putting
%==========================================================
\begin{eqnarray}
\label{HOMSPHE1}
x_\mu  = R \frac{u_\mu }{u_3} =
\frac{u_\mu }{\sqrt{1- (u_1^2+u_2^2)/R^2}},
\,\,\,\, \mu = 1,2.
\end{eqnarray}
%==========================================================
Geometrically ($x_1, x_2$) correspond to a projection from the center
of the sphere to a tangent plane at the North pole.
In this parametrization the metric tensor has the following form
%===================================================================
\begin{eqnarray}
g_{\mu\nu}=
\frac{1}{1+ {r^2}/{R^2}}
\left[\delta_{\mu\nu} + \frac{x_\mu x_\nu}{r^2}
\frac{1}{1+ {r^2}/{R^2}} \right],
\,\,\,\,\,\,
r^2 = x_\mu x_\mu,
\end{eqnarray}
%====================================================================
The Laplace-Beltrami operator corresponds to
%====================================================================
\begin{eqnarray}
\label{LAPLACE2}
\Delta_{LB}
&=&\left(1 + \frac{r^2}{R^2}\right)
\left[\frac{\partial^2}{\partial x_{\mu}^2}
+ \frac{x_\mu}{R^2}\frac{\partial}{\partial x_{\mu}}
+ \frac{1}{R^2}\left(x_\mu \frac{\partial}{\partial x_{\mu}}\right)^2
\right]
=
\left(\pi_1^2 + \pi_2^2 + \frac{L_3^2}{R^2} \right),
\end{eqnarray}
%=====================================================================
where
%=====================================================================
\begin{eqnarray}
\pi_\mu
= \left(\frac{\partial}{\partial x_{\mu}}
+ \frac{x_\mu x_\nu}{R^2}\frac{\partial}{\partial x_{\nu}}
\right),
\,\,\,\,\,
L_{3}
= \left(x_1 \frac{\partial}{\partial x_{2}}
- x_2 \frac{\partial}{\partial x_{1}}\right).
\end{eqnarray}
%===========================================================
Using the connection between operators $\pi_\mu$ and the generators
of the $O(3)$ group
%==========================================================
$$
- \frac{L_1}{R} = \pi_2
\qquad
\frac{L_2}{R} =
\pi_1,
\qquad
L_3 = -(x_1 \pi_2 - x_2 \pi_1)
$$
%==========================================================
we obtain the following commutation relations
%===========================================================
\begin{eqnarray}
[L_3, \pi_1] =  \pi_2,
\,\,\,\,\,\,\,\,\,
[L_3, \pi_2] =  - \pi_1,
\,\,\,\,\,\,\,\,
[\pi_1, \pi_2] = \frac{L_3}{R^2},
\end{eqnarray}
%===========================================================
so that for $R\rightarrow \infty$ the o(3) algebra contracts
to the e(2) one. Moreover the momenta $\pi_\mu $ contract to
$P_\mu =  \partial/\partial x_\mu$, ($\mu = 1,2$) and the
o(3) Laplace-Beltrami operator (\ref{LAPLACE2}) contracts to the
e(2) one:
%===========================================================
\begin{eqnarray}
\Delta_{LB} = \pi_1^2 + \pi_2^2 + \frac{L_3^2}{R^2}
\rightarrow \Delta = (P_1^2 + P_2^2).
\end{eqnarray}
%============================================================

\subsection{Contractions from $o(2,1)$ to $e(2)$}

As in section 3.1, let us introduce the Beltrami coordinates on
the hyperboloid $H_2$ putting
%==================================================================
\begin{eqnarray}
\label{BELT2}
  x_{\mu} = R\frac{u_{\mu}}{u_0} = R\frac{u_{\mu}}{\sqrt{R^2+
  u_1^2 + u_2^2}},\,\,\,\,\,   \mu = 1,2.
\end{eqnarray}
%==================================================================
The $O(2,1)$ generators can be expressed as:
%==================================================================
\begin{eqnarray*}
 - \frac{K_1}{R} \equiv {\tilde \pi}_2  = \frac{\partial}{\partial x_2} -
 \frac{x_2}{R^2} \left(x_1 \frac{\partial}{\partial x_1}
+ x_2 \frac{\partial}{\partial x_2}\right),
\qquad
- \frac{K_2}{R} \equiv {\tilde \pi}_1  = p_1 -
\frac{x_1}{R^2} \left(x_1 \frac{\partial}{\partial x_1}
+ x_2 \frac{\partial}{\partial x_2}\right),
\end{eqnarray*}
$$
L_3 =  x_1 \pi_2 - x_2 \pi_1.
$$
%==================================================================
The commutation relations of the o(2,1) algebra (\ref{NONCOMPACT1})
in terms of the new operators take the form
%==================================================================
\begin{equation}
\left[{\tilde \pi}_1, {\tilde \pi}_2\right]
=   - \frac{L_3}{R^2},\,\,\,\,
\left[L_3, {\tilde \pi}_1\right]   =     {\tilde \pi}_2,\,\,\,\,
\left[{\tilde \pi}_2, L_3 \right]   =    {\tilde \pi}_1,
\end{equation}
%==================================================================
so, that for $R\rightarrow\infty$ the o(2,1) algebra contracts
to e(2) and the momenta ${\tilde \pi}_\mu$ to $P_\mu =
\partial/\partial x_\mu$. The o(2,1) Laplace-Beltrami operator
(2.2) contracts to the e(2) one:
%===========================================================
\begin{eqnarray}
\Delta_{LB} = {\tilde \pi}_1^2 + {\tilde \pi}_2^2 - \frac{L_3^2}{R^2}
\rightarrow  \Delta = (P_1^2 + P_2^2).
\end{eqnarray}
%============================================================

\subsection{Contractions from $o(2,1)$ to $e(1,1)$}

Let us introduce Beltrami coordinates on hyperboloid $H_2$
%==================================================================
\begin{eqnarray}
\label{BELT3}
y_0 = R\frac{u_{0}}{u_2} = R\frac{u_{0}}{\sqrt{u_0^2+u_1^2-R^2}},
 \,\,\,\,\,
y_1 = R\frac{u_{1}}{u_2} = R\frac{u_{1}}{\sqrt{u_0^2+u_1^2-R^2}}.
\end{eqnarray}
%==================================================================
The $O(2,1)$ generators can be expressed as
%==================================================================
$$
- \frac{K_1}{R} \equiv {\tilde {\tilde \pi}}_1  =
\frac{\partial}{\partial y_0} -  \frac{y_0}{R^2}
\left(y_0 \frac{\partial}{\partial y_0} +
y_1 \frac{\partial}{\partial y_1}\right),
\,\,\,
- \frac{L_3}{R} \equiv {\tilde {\tilde \pi}}_2
= \frac{\partial}{\partial y_1} +  \frac{y_1}{R^2}
\left(y_0 \frac{\partial}{\partial y_0}
+ y_1 \frac{\partial}{\partial y_1}\right),
$$
\begin{eqnarray}
- K_2 \equiv  K = y_0 {\tilde {\tilde \pi}}_2 +
y_1 {\tilde {\tilde \pi}}_1.
\end{eqnarray}
%==================================================================
The commutators of the $o(2,1)$ algebra (\ref{NONCOMPACT1}) in the new
operators (${\tilde {\tilde \pi}}_1, {\tilde {\tilde \pi}}_2, K$)
take the form
%==================================================================
\begin{equation}
\left[{\tilde {\tilde \pi}}_1, {\tilde {\tilde \pi}}_2\right]
=   \frac{K}{R^2},\,\,\,\,
\left[K, {\tilde {\tilde \pi}}_1\right]
=   -  {\tilde {\tilde \pi}}_2,\,\,\,\,
\left[{\tilde {\tilde \pi}}_2, K \right]
=  {\tilde {\tilde \pi}}_1.
\end{equation}
%==================================================================
so, that for $R\rightarrow\infty$ the $o(2,1)$ algebra contracts
to the $e(1,1)$ one. The $o(2,1)$ Laplace-Beltrami operator contracts
to the $e(1,1)$ one:
%===========================================================
\begin{eqnarray}
\Delta_{LB} = {\tilde {\tilde \pi}}_1^2
- {\tilde {\tilde  \pi}}_2^2 + \frac{K^2}{R^2}
\rightarrow
\frac{\partial^2}{\partial y_0^2} - \frac{\partial^2}{\partial y_1^2},
\end{eqnarray}
%============================================================
and  eq.~(\ref{HE1}) transforms for large $\ell\sim pR$ to the
one-dimensional Klein-Gordan equation.
%===========================================================
\begin{eqnarray}
\frac{\partial^2\psi}{\partial y_0^2}
-
\frac{\partial^2\psi}{\partial y_1^2}
+ p^2 \psi = 0.
\end{eqnarray}
%============================================================

\section{Contraction for systems of coordinates}

\subsection{Contractions and coordinate systems on $S_2$}

\vspace{0.3cm}
{\bf 1. Spherical coordinates on $S_2$ to polar on $E_2$}.

We consider the spherical coordinate (\ref{COOR1}) with the parametr
$a_1=a_2$ and put
%============================================================
\begin{eqnarray*}
\tan \theta = \frac{r}{R}
\end{eqnarray*}
%============================================================
In the contraction limit $R\to\infty$, $\theta\to 0$ we have
%============================================================
\begin{eqnarray*}
x_1 = R \frac{u_1}{u_3}
\rightarrow x = r\cos\vphi,
\qquad
x_2 = R \frac{u_2}{u_3} \rightarrow y = r\sin\vphi
\end{eqnarray*}
%============================================================
and
%============================================================
\begin{eqnarray*}
Q_S =  L_3^2 \rightarrow L^2
\end{eqnarray*}
%============================================================

\vspace{0.3cm}
\noindent
{\bf 2. Spherical coordinate on $S_2$ to Cartesian on $E_2$}.

We choose the spherical coordinate (\ref{COOR1}) with $a_2=a_3$.
Putting
%==============================================================
\begin{eqnarray*}
\cos\theta' \sim \frac{x}{R} \sim 0,
\qquad
\cos\vphi' \sim \frac{y}{R} \sim 0,
\end{eqnarray*}
%==============================================================
and taking the limit $R\rightarrow\infty$ and
$\theta'\rightarrow\frac{\pi}{2}, \, \vphi'\rightarrow\frac{\pi}{2}$,
we obtain
%============================================================
\begin{eqnarray*}
\frac{1}{R^2} Q_S = \frac{L_1^2}{R^2} = \pi_1^2
\rightarrow P_1^2 \sim Q_C
\end{eqnarray*}
%============================================================
and
%============================================================
\begin{eqnarray*}
x_1 = R \, \frac{\cot\theta'}{\sin\vphi'} \rightarrow x
\,\,\,\,\,\,\,
x_2 = R \, \cot\vphi' \rightarrow y.
\end{eqnarray*}
%============================================================
It is easy to see that for the case $a_1=a_3$ the corresponding
spherical system of coordinates (\ref{COOR1}) contracts to Cartesian
coordinates on $E_2$ for $R\rightarrow\infty$.

\vspace{0.3cm}
\noindent
{\bf 3. Elliptic coordinates on $S_2$ to elliptic coordinates on $E_2$.}

We take $Q$ in its general form, equivalent to
%============================================================
\begin{eqnarray}
Q_E = L_3^2 -
\left(\frac{a_2-a_1}{a_3-a_2}\right)
L_1^2.
\end{eqnarray}
%============================================================
We put
%============================================================
\begin{eqnarray}
\label{ELLIP1}
\frac{R^2}{a_3-a_1} = \frac{D^2}{a_2-a_1},
\end{eqnarray}
%============================================================
and in the limit $R^2 \sim  a_3 \rightarrow \infty$ obtain
%============================================================
\begin{eqnarray}
Q_E =L_3^2 - \left(\frac{a_3-a_1}{a_3-a_2}\right) \frac{D^2}{R^2}
\, L_1^2\, \rightarrow L_3^2 - D^2 P_2^2 \sim I_E.
\end{eqnarray}
%============================================================
For the coordinates we put
%=============================================================
\begin{eqnarray}
\label{ELLIP-TRIG1}
\rho_1 = a_1 + (a_2 - a_1)\cos^2\eta,\,\,\,\,\,\,\,\,\,
\rho_2 = a_1 + (a_2 - a_1)\cosh^2\xi,
\end{eqnarray}
%==============================================================
and for $R^2\sim a_3 \rightarrow\infty$, using eq.~(\ref{ELLIP1}),
we obtain eq.(\ref{ELLTRIG1}), i.e. elliptic coordinates on the plane
$E_2$.

\vspace{0.3cm}
\noindent
{\bf 4. Elliptic coordinates on $S_2$ to Cartesian coordinates
on $E_2$.}

We start from the coordinates (\ref{ALGEB1}) but change the ordering
of the parameters $a_i$, which corresponds the interchange of coordinates
$u_3 \leftrightarrow u_2$, i.e. put
%==============================================================
\begin{eqnarray}
\label{CART-ELLIP0}
a_1 \leq \rho_1 \leq a_3 \leq \rho_2 \leq a_2,
\end{eqnarray}
%==============================================================
and choose $a_3-a_1 = a_2-a_3 \equiv a$. Than we have
%============================================================
\begin{eqnarray}
Q_{E} = a(L_2^2-L_1^2).
\end{eqnarray}
%============================================================
Introducing the new coordinates by
%==============================================================
\begin{eqnarray}
\label{CART-ELLIP1}
\frac{a_3 - \rho_1}{a} = \xi_1 \,\,\,\,\,\,\,\,
\frac{\rho_2 - a_3}{a} = \xi_2,
\end{eqnarray}
%==============================================================
we can rewrite the (\ref{ALGEB1}) in the form
%==============================================================
\begin{eqnarray}
\label{CARELL1}
u_1^2 = \frac{R^2}{2} (1-\xi_1)(1+\xi_2),
\,\,\,\,\,\,\,
u_2^2 = \frac{R^2}{2} (1+\xi_1)(1-\xi_2),
\,\,\,\,\,\,\,
u_2^2 = R^2 \xi_1\xi_2.
\end{eqnarray}
%==============================================================
Using eq.~(\ref{HOMSPHE1}) we have for Beltrami coordinates
%==============================================================
\begin{eqnarray}
\label{CARELL11}
x_1^2 = R^2 \frac{(1-\xi_1)(1+\xi_2)}{2\xi_1\xi_2}
\,\,\,\,\,\,\,\,
x_2^2 = R^2 \frac{(1+\xi_1)(1-\xi_2)}{2\xi_1\xi_2}.
\end{eqnarray}
%==============================================================
>From equation (\ref{CARELL11}) we obtain
%==============================================================
\begin{eqnarray}
\label{CARELL111}
\xi_{2,1} = \frac{R^2}{R^2+x_1^2+x_2^2}
\left\{\left[1+\frac{x_1^2+x_2^2}{R^2} +
\frac{(x_1^2-x_2^2)^2}{4R^4}\right]^{1/2}
\mp \frac{x_1^2-x_2^2}{2R^2}\right\}.
\end{eqnarray}
%==============================================================
Taking now the limit $R\rightarrow \infty$ we have
%==============================================================
\begin{eqnarray}
\label{CART-ELLIP2}
\xi_{1} \rightarrow 1 - \frac{x^2}{R^2}
\,\,\,\,\,\,\,\,\,\,\,
\xi_{2} \rightarrow 1 - \frac{y^2}{R^2},
\end{eqnarray}
%==============================================================
and hence $x_1$ and $x_2$ of eq.~(\ref{CARELL11}) go into Cartesian
coordinates:
%==============================================================
\begin{eqnarray}
x_{1} \rightarrow x
\,\,\,\,\,\,\,\,\,\,\,
x_{2} \rightarrow y.
\end{eqnarray}
%==============================================================
For the integral of motion in the limit $R^2 \sim a \rightarrow \infty$
we have
%============================================================
\begin{eqnarray}
\frac{1}{a R^2} Q_{E} = (\pi_1^2-\pi_2^2) \rightarrow P_1^2 - P_2^2
= Q_{C}.
\end{eqnarray}
%============================================================

\vspace{0.3cm}
\noindent
{\bf 5. Elliptic coordinates on $S_2$ to parabolic coordinates on
$E_2$.}

We take the operator (\ref{ALGEB1}) with
$a_1 \leq \rho_1 \leq a_2 \leq \rho_2 \leq a_3$ and choose the parameter
$a_3-a_2 = a_2-a_1 \equiv a$. We must first "undo"  the diagonalization
(\ref{INT-MOTION7}) by a rotation through $\pi/4$.
The operator (\ref{ENALGEB1}) transforms into
%============================================================
\begin{eqnarray}
\label{ELROTOP1}
\frac{1}{aR}Q_{E} = - \frac{1}{R} (L_1 L_3 + L_3 L_1) =
(L_3 \pi_2 + \pi_2 L_3),
\end{eqnarray}
%============================================================
with the correct limit (\ref{PAR2}) for $R\rightarrow\infty$. The
coordinates (\ref{EL1}) on $S_2$ are rotated into
%=============================================================
\begin{eqnarray}
\label{ELROT2}
u_1 = \frac{R}{\sqrt 2} (\sn\alpha \dn\beta + \dn\alpha \sn\beta),
\,\,\,\,\,\,
u_2 = R \cn\alpha \cn\beta,
\,\,\,\,\,\,
u_3 = \frac{R}{\sqrt 2} (\dn\alpha \sn\beta -
\sn\alpha \dn\beta),
\end{eqnarray}
%=================================================================
with modulus $k=k'=1/\sqrt 2$ for all Jacobi elliptic functions.

>From eq~(\ref{ELROT2}) we obtain
%====================================================================
\begin{eqnarray}
\label{ELLIM1}
\sn\alpha &=& \frac{1}{\sqrt 2}\left[
\left(1+\frac{u_1}{R}\right)^{1/2}
\left(1-\frac{u_3}{R}\right)^{1/2}
-
\left(1-\frac{u_1}{R}\right)^{1/2}
\left(1+\frac{u_3}{R}\right)^{1/2}
\right]
\nonumber\\[2mm]
{\sqrt 2}
\dn\beta &=& \frac{1}{\sqrt 2}\left[
\left(1+\frac{u_1}{R}\right)^{1/2}
\left(1-\frac{u_3}{R}\right)^{1/2}
+
\left(1-\frac{u_1}{R}\right)^{1/2}
\left(1+\frac{u_3}{R}\right)^{1/2}
\right].
\end{eqnarray}
%=================================================================
Eq.~(\ref{ELLIM1}) suggest the limiting procedure. Indeed we put
%====================================================================
\begin{eqnarray}
\label{ELLIM000}
\sn\alpha = -1 + \frac{u^2}{2R}, \,\,\,\,\,\,\,
{\sqrt 2}
\dn\beta = 1 + \frac{v^2}{2R}.
\end{eqnarray}
%=================================================================
In the limit $R\rightarrow\infty$ we obtain
%==============================================================
\begin{eqnarray}
x_{1} \rightarrow x = \frac{u^2-v^2}{2},
\,\,\,\,\,\,\,\,\,\,\,
x_{2} \rightarrow y = u v,
\end{eqnarray}
%==============================================================
i.e. the parabolic coordinates (\ref{PAR2}).

\subsection{Contractions of coordinate systems from $H_2$ to $E_2$}

\vspace{0.3cm}
\noindent
{\bf 1. Pseudo-spherical Coordinates on $H_2$ to Polar
Coordinates on $E_2$}

In the limit $R\rightarrow\infty$, $\tau\rightarrow 0$
putting $\tanh\tau \sim r/R$ we have:
%============================================================
\begin{eqnarray*}
{Q}_{S} = L_3^2 \rightarrow L_3^2
\end{eqnarray*}
%============================================================
and for Beltrami coordinates (\ref{BELT2}) we obtain:
%============================================================
\begin{eqnarray*}
x_1 = R \frac{u_1}{u_0}
\rightarrow x = r\cos\varphi,\,\,\,\,\,
x_2 = R \frac{u_2}{u_0} \rightarrow y = r\sin\varphi.
\end{eqnarray*}
%============================================================

\vspace{0.3cm}
\noindent
{\bf 2. Equidistant Coordinates on $H_2$ to Cartesian on $E_2$}

For Beltrami coordinates (\ref{BELT2}) we have:
%==============================================================
\begin{eqnarray}
\label{BELT-PS}
x_1 = R\tanh\tau_2,\,\,\,\,\,  x_2 = R\tanh\tau_1/\cosh\tau_2.
\end{eqnarray}
%==============================================================
Taking the limit $R\rightarrow\infty$, $\tau_1, \tau_2 \rightarrow 0$
and putting $\sinh\tau_1 \sim y/R,\,\,\, \sinh\tau_2 \sim x/R$ in
(\ref{BELT-PS}) we obtain $x_1\rightarrow x,\,\,\,  x_2\rightarrow y$
and
%============================================================
\begin{eqnarray*}
\frac{{Q}_{EQ}}{R^2} = \pi_1^2 \rightarrow  P_1^2 \sim Q_{C}.
\end{eqnarray*}
%============================================================

\vspace{0.3cm}
\noindent
{\bf 3. Horocyclic Coordinates on $H_2$ to Cartesian on $E_2$}

For variables $\tilde x, \tilde y$ we obtain:
%============================================================
\begin{eqnarray*}
  \tilde x = \frac{u_2}{u_0 - u_1},\,\,\,\,\, \tilde y =
  \frac{R}{u_0 - u_1}.
\end{eqnarray*}
%============================================================
In the limit $R\rightarrow\infty$ we get: $\tilde x \rightarrow
y/R,\,\,\,\,\, \tilde y \rightarrow 1 + x/R$
and Beltrami coordinates go into Cartesian ones
%============================================================
\begin{eqnarray*}
x_1 = R\, \frac{\tilde x^2+\tilde y^2-1}{\tilde x^2+\tilde y^2+1}
\to x,
\qquad
x_2 = \frac{2 \tilde x R}{\tilde x^2+\tilde y^2+1}
\to y.
\end{eqnarray*}
%============================================================
For integral of motion we have:
%============================================================
\begin{eqnarray*}
\frac{{Q}_{HO}}{R^2} = \pi_2^2 + \frac{L_3^2}{R^2} -
\frac{1}{R}\{\pi_2, L_3\} \rightarrow  P_2^2 \sim Q_{C}.
\end{eqnarray*}
%============================================================

\vspace{0.3cm}
\noindent
{\bf 4. Elliptic Coordinates on $H_2$ to Elliptic Coordinates on $E_2$}

We put
%============================================================
\begin{eqnarray}
\label{RAD1}
\frac{R^2}{a_2 - a_3} = \frac{D^2}{a_1 -a_2}.
\end{eqnarray}
%============================================================
and in the limit $R^2 \sim (-a_3) \rightarrow \infty$ obtain:
%============================================================
\begin{eqnarray*}
 {Q}_E = L_3^2 + \frac{D^2}{R^2} K_2^2
  \rightarrow L^2 + D^2p_1^2 \sim Q_{E},
\end{eqnarray*}
%============================================================
where $2D$ is the focal distance. Writting the coordinates as
%============================================================
\begin{eqnarray*}
\rho_1 = a_1 + (a_1-a_2) \sinh^2\xi,
\,\,\,\,\,\,\,\,
\rho_2 = a_2 + (a_1-a_2) \cos^2\eta
\end{eqnarray*}
%============================================================
and using eq.(\ref{RAD1}) in the limit $R^2 \sim (- a_3)
\rightarrow \infty$  we get the ordinary elliptic coordinates
on $E_2$ plane \cite{WLS,MILLER1}

\vspace{0.3cm}
\noindent
{\bf 5. Elliptic coordinates on $H_2$ to Cartesian on $E_2$}.

We make a special choice of the parameters $a_i$: $a_1-a_2 = a_2-a_3$
and determine new variables $\xi_{1,2}$ by the formula
%============================================================
\begin{eqnarray}
\label{HYPCAR1}
\xi_{1,2} = \frac{\vrho_{1,2}-a_2}{a_1-a_2} =
\frac{u_0^2+u_2^2}{2R^2} \pm
\sqrt{\left(\frac{u_0^2+u_2^2}{2R^2}\right)^2
- \frac{u_1^2}{R^2}},
\end{eqnarray}
%============================================================
Considering the limit $R\to\infty$ we obtain: $\xi_1 \sim 1 + 2 y^2/R^2$,
$\xi_2 \sim x^2/R^2$ and the Beltrami coordinate (\ref{BELT2})
take the Cartesian form
%============================================================
\begin{eqnarray*}
\label{ELLIP-CAR2}
x_1 = R \frac{u_1}{u_0} = R
\sqrt{\frac{2 \xi_1 \xi_2}{(\xi_1+1)(\xi_2+1)}}
\rightarrow  \, x,
\,\,\,\,\,
x_2 = R \frac{u_2}{u_0} = R
\sqrt{\frac{(\xi_1-1)(1-\xi_2)}
{(\xi_1+1)(\xi_2+1)}} \rightarrow  \, y
\end{eqnarray*}
%============================================================
The operator ${Q}_E$ goes to Cartesian one
%============================================================
\begin{eqnarray*}
\frac{{Q}_E}{R^2} = \frac{L_3^2}{R^2} + \pi_1^2
\rightarrow P_1^2 \sim Q_C.
\end{eqnarray*}
%============================================================

\vspace{0.3cm}
\noindent
{\bf 6. Elliptic coordinates on $H_2$ to parabolic on $E_2$}.

We start from the rotated elliptic coordinates
%==============================================================
\begin{equation}
\label{ROT-ELLIP10}
\left(\begin{array}{c} u_0' \\ u_1' \\ u_2'\end{array}\right)
=\left(\begin{array}{ccc}
\cosh f &\sinh f &0\\ \sinh f &\cosh f &0\\ 0 &0 &1\end{array}\right)
\left(\begin{array}{c} u_0 \\ u_1 \\ u_2\end{array}\right)
=\left(\begin{array}{c} u_0\cosh f+u_1\sinh f \\
u_0\sinh f+u_1\cosh f \\ u_2\end{array}\right),
\end{equation}
%==============================================================
where $\sinh^2 f = (a_1-a_2)/(a_2-a_3)$.
We choose $a_2-a_3=a_1-a_2\equiv a$. Then for rotated elliptic coordiantes
(\ref{ROT-ELLIP10}) we get
%==============================================================
\begin{eqnarray}
\label{ROT-ELLIP1}
u_0 = {R\over \sqrt{2}} (\sn \alpha \dn \beta\ +
i\sqrt{2} \cn \alpha \cn \beta ), \,\,
u_1 = {R \over \sqrt{2}} (i \cn \alpha \cn \beta
+ \sqrt{2} \sn \alpha \dn \beta),
\,\,
u_2 = i R \dn \alpha \sn \beta,
\end{eqnarray}
%==============================================================
with modulus $k=k'=1/\sqrt{2}$ for all Jacobi elliptic function.
The integral of motion transforms into
%==============================================================
\begin{equation}
{Q}_{E'}= 3 \, L_3^2-  \sqrt{2} \, (K_1 L_3 + L_3 K_1),
\end{equation}
%==============================================================
with the correct limit to (\ref{PAR1}). From eq. (\ref{ROT-ELLIP1})
we obtain
%============================================================
\begin{eqnarray*}
\cn\alpha &=& -\frac{i}{2}\sqrt{\left(1 + \frac{u'_1}{R\sqrt{2}} -
\frac{u'_0}{R}\right)^2 + \frac{u'^2_2}{2R^2}} +
\frac{i}{2}\sqrt{\left(1 - \frac{u'_1}{R\sqrt{2}} + \frac{u'_0}{R}
\right)^2 + \frac{u'^2_2}{2R^2}},
\\[2mm]
\cn\beta &=& \frac{1}{2}\sqrt{\left(1 + \frac{u'_1}{R\sqrt{2}} -
\frac{u'_0}{R}\right)^2 + \frac{u'^2_2}{2R^2}} +
\frac{1}{2}\sqrt{\left(1 - \frac{u'_1}{R\sqrt{2}} + \frac{u'_0}{R}
\right)^2 + \frac{u'^2_2}{2R^2}},
\end{eqnarray*}
%============================================================
and therefore for large $R$ we have
%============================================================
\begin{eqnarray*}
-i \cn\alpha \simeq 1 - \frac{1}{2\sqrt{2}}\frac{u^2}{R},
\,\,\,\,\,\,\,
\cn\beta \simeq 1 + \frac{1}{2\sqrt{2}}\frac{v^2}{R}.
\end{eqnarray*}
%============================================================
In the limit $R\rightarrow\infty$ we obtain
%============================================================
\begin{eqnarray*}
x_1 \rightarrow x = \frac{u^2-v^2}{2},
\,\,\,\,\,
x_2 \rightarrow y = uv,
\end{eqnarray*}
%============================================================
i.e. the parabolic coordinates (\ref{PAR2}).

\subsection{Contractions of coordinate systems from $H_2$ to $E_{1,1}$}

\vspace{0.3cm}
\noindent
{\bf 1. Equidistant coordinates on $H_2$ to pseudo-spherical ones
on $E_{1,1}$ plane}.

For Beltrami coordinates (\ref{BELT3}) we have:
%==============================================================
\begin{eqnarray}
  y_0 = R\coth\tau_1\cosh\tau_2,
\,\,\,\,\,
  y_1 = R\coth\tau_1\sinh\tau_2.
\end{eqnarray}
%==============================================================
Taking the limit $R\rightarrow\infty$, $\tau_1\rightarrow
i\frac{\pi}{2}+ \frac{r}{R}$ and putting
%==============================================================
\begin{eqnarray}
\coth\tau_1 = \tanh\frac{r}{R} \sim \frac{r}{R},
\end{eqnarray}
%==============================================================
we obtain
%============================================================
\begin{eqnarray}
  y_0 \rightarrow t = r \cosh\tau_2,
\,\,\,\,\,
  y_1 \rightarrow x = r \sinh\tau_2,
\end{eqnarray}
%============================================================
where $0 \leq r < \infty$, $ -\infty < \tau_2 < \infty$.
For the integral of motion we get
%============================================================
\begin{eqnarray}
{Q}_{EQ} = K_2^2 \rightarrow Q_{S} = L_3^2.
\end{eqnarray}
%============================================================

\vspace{0.3cm}
\noindent
{\bf 2. Pseudo-spherical coordinates on $H_2$ to Cartesian
coordinates on $E_{1,1}$}

For coordinates (\ref{BELT3}) we have
%==============================================================
\begin{eqnarray}
  y_0 = R\frac{\coth\tau}{\cos\varphi},
\,\,\,\,\,
  y_1 = R\cot\varphi.
\end{eqnarray}
%==============================================================
Taking the limit $R\rightarrow\infty$, $\tau\rightarrow i\pi/2$,
$\varphi \rightarrow \frac{\pi}{2}$ and putting
%==============================================================
\begin{eqnarray}
\coth\tau \sim \frac{t}{R},\,\,\,\
\cot\varphi \sim \frac{x}{R},
\end{eqnarray}
%==============================================================
we see that Beltrami coordinates go into Cartesian ones
%============================================================
\begin{eqnarray}
  y_0 \rightarrow t,
\,\,\,\,\,
  y_1 \rightarrow x.
\end{eqnarray}
%============================================================
For the integral of motion we obtain
%============================================================
\begin{eqnarray}
\frac{{Q}_{S}}{R^2} = \frac{L_3^2}{R^2} \rightarrow  P_2^2
\sim {Q}_{C}.
\end{eqnarray}
%============================================================

\section{Contraction of basis functions on $S_2$ and $H_2$}

\subsection{Contraction of spherical basis and interbasis expansions}

{\bf 5.1.1. Spherical basis on $S_2$ to polar on $E_2$}.

We start from the standard spherical functions $Y_{lm}(\theta, \phi)$
as basis functions of irreducible representations of the group O(3)
(see e.g.\ ref.~\cite{VAR})
%==============================================================
\begin{eqnarray}
Y_{lm}(\theta, \phi)
&=& (-1)^{\frac{m+|m|}{2}}
\left[\frac{2l+1}{2} \frac{(l+|m|)!}{(l-|m|)!} \right]^{1/2}
\frac{(\sin\theta)^{|m|}}{2^{|m|} |m|!}
\nonumber\\[2mm]
&\cdot &_2F_1\left(-l+|m|, l+|m|+1; |m|+1; \sin^2\frac{\theta}{2}\right)
\,
\frac{e^{im\phi}}{\sqrt{2\pi}}
\end{eqnarray}
%==============================================================
In the contraction limit $R\rightarrow\infty$ we put
%==============================================================
\begin{eqnarray}
\tan\theta \sim \theta \sim \frac{r}{R},
\,\,\,\,\,\,\,
l \sim k R.
\end{eqnarray}
%==============================================================
Using the asymptotic formulas
%==============================================================
\begin{eqnarray}
\label{GAMMA-FUN10}
\lim_{R\rightarrow\infty}
{_2F_1}\left(-kR, kR; |m|+1; \frac{r^2}{4R^2}\right)
&=& {_0F_1}\left(|m|+1; -\frac{k^2 r^2}{4}\right)
\\[2mm]
\label{GAMMA-FUN1}
\lim_{z\rightarrow\infty} \frac{\Gamma(z+\alpha)}{\Gamma(z+\beta)}
&=& z^{\alpha-\beta}
\end{eqnarray}
%==============================================================
and formula
%==============================================================
\begin{eqnarray}
J_{\nu}(z)
= \left(\frac{z}{2}\right)^{\nu}\frac{1}{\Gamma(\nu+1)}
{_{0}F_{1}}
\left(\nu+1; - \frac{z^{2}}{4}\right)
\end{eqnarray}
%==============================================================
we obtain
%==============================================================
\begin{eqnarray}
\label{BESSEL1}
\lim_{\scriptstyle R\rightarrow\infty\atop\scriptstyle
\theta\rightarrow 0}
\frac{1}{\sqrt R} Y_{lm}(\theta, \phi)
= (-1)^{\frac{m+|m|}{2}} \sqrt{k} J_{|m|}(kr)
\frac{e^{im\phi}}{\sqrt{2\pi}}
\end{eqnarray}
%==============================================================
The result (\ref{BESSEL1}) is not new \cite{VAR}. The point is that
this asymptotic formula is obtained very naturally in the
context of group contractions applied to the separation of
variables.

\vspace{0.3cm}
\noindent
{\bf 5.1.2. Spherical basis on $S_2$ to Cartesian on $E_2$}

We start from the coordinates ($\theta', \phi'$) in eq. (\ref{COOR1}),
but drop the primes, and write the corresponding spherical functions
as
%==============================================================
\begin{eqnarray}
Y_{lm}(\theta, \phi)
&=&
\frac{\sqrt{2l+1}}{2\pi}
e^{im\phi}
(\sin\theta)^{|m|}
\nonumber\\[2mm]
&&\times
\cases{(-1)^{\frac{l+m}{2}}\left[\frac{\Gamma(\frac{l+m+1}{2})
\Gamma(\frac{l-m+1}{2})}{\Gamma((\frac{l+m+2}{2})
\Gamma(\frac{l-m+2}{2})}
\right]^{\frac{1}{2}} \,
{_2F_1}\left(-\frac{l-m}{2},
\frac{l+m+1}{2};\frac{1}{2};\cos^{2}\theta\right)
\cr
\cr
\cr
(-1)^{\frac{l+m-1}{2}}
\big[ \frac{\Gamma(\frac{l+m+2}{2})
\Gamma(\frac{l-m+2}{2})}{\Gamma((\frac{l+m+1}{2})
\Gamma(\frac{l-m+1}{2})}
\big]^{\frac{1}{2}} \, 2\cos\theta \,
{_2F_1}\left(-\frac{l-m-1}{2},
\frac{l+m+2}{2};\frac{3}{2};\cos^{2}\theta\right)
\cr}
\end{eqnarray}
%==============================================================
for $l+m$ even and odd, respectively, we now put
%===============================================================
\begin{eqnarray}
l \sim kR, \qquad m \sim k_{2}R, \qquad
\theta \sim\frac{\pi}{2},\qquad
\phi \sim \frac{\pi}{2},
\end{eqnarray}
%===============================================================
and
%===============================================================
\begin{eqnarray}
\sin\theta \rightarrow  1,
\,\,\,\,\, \cos\theta \rightarrow \frac{x}{R},
\,\,\,\,\,\,\,\,
\cos \phi \rightarrow \frac{y}{R}.
\end{eqnarray}
%===============================================================
The $_{2}F_{1}$ hypergeometric functions simplify to $_{0}F_{1}$ ones,
the $\Gamma$ functions also simplify and the final result is
that under the contraction we have
%================================================================
\begin{eqnarray}
\label{SPHER-CARTES0}
\lim_{R\rightarrow\infty}
(-1)^{-\frac{l+m}{2}}
\,
Y_{lm}(\theta, \phi)
&=&
\sqrt{\frac{k}{k_1}}
\cdot
\frac{e^{ik_{2}y}}{\pi}
\cdot
\left\{\matrix{
{_0F_1}\left(\frac{1}{2}; \frac{-k_1^2x^2}{4}\right)
\cr
\cr
{-i}(k_1 x)
{_{0}F_{1}} \left(\frac{3}{2}; \frac{{-k_{1}}^{2}x^{2}}{4}\right)
\cr}\right.
\nonumber\\[2mm]
&=&
\label{SPHETCAR1}
\sqrt{\frac{k}{k_1}}
\cdot
\frac{e^{ik_{2}y}}{\sqrt{\pi}}
\cdot
\left\{\matrix{\cos{k}_{1}x
\cr
\cr
- i\sin{k}_{1}x \cr}
\right.
\end{eqnarray}
%===============================================================
with ${k_{1}}^{2}+{k_{2}}^{2}=k^{2}$ and for $l+m$ even and odd,
respectively. The parity properties of $Y_{lm}$ under the exchange
$\theta \rightarrow \pi-\theta$ have lead to the appearance of
$\cos{k}_{1}x$ and $\sin{k}_{1}x$ in eq.(\ref{SPHETCAR1}), instead
of the usual
Cartesian coordinate solution $\exp i(k_{1}x+k_{2}y)$.

Finally note that, the factor $\sqrt{\frac{k}{k_1}}$ in formula
(\ref{SPHETCAR1}) are connected with the contraction of Kronecker
symbols to delta function
%===============================================================
\begin{eqnarray*}
\label{DELTA1}
\delta_{ll'} \to \frac{1}{R} \delta (k-k') = \frac{1}{R} \frac{k}{k'}
\delta (k_1-k'_1)
\end{eqnarray*}
%===============================================================

\vspace{0.3cm}
\noindent
{\bf 5.1.3. Contraction in interbasis expansions}.

Let us now to consider the contraction $R\rightarrow\infty$ for the
interbasis expansion (\ref{EXP1}). The contraction of basis functions
were presented in the formulas  (\ref{BESSEL1}) and (\ref{SPHETCAR1}).
In order to obtain the corresponding limit we need the asymptotic
behavior of the "little" Wigner  $d$-function for a large $R$.
It is easy to see that the expression of $d$-function for the angle
$\pi/2$ in terms of hypergeometric functions $_2F_1$ (see for example
\cite{VAR}) is not applicable for the contraction limit when
$\ell\to\infty$
and $m\to \infty$ simultaneously. To make this contraction we use an
integral representation for the function $d^l_{m_2, m_1}(\beta)$
\cite{VAR}
%================================================================
\begin{eqnarray}
\label{INT-DFUN0}
d^l_{m_2, m_1}(\beta)
&=& \frac{i^{m_2-m_1}}{2\pi}
\,\,
\left[\frac{(l+m_2)!(l-m_2)!}{(l+m_1)!(l-m_1)!}\right]^{\frac12}
\\[2mm]
&\times&
\int_0^{2\pi}
\left(e^{i\frac{\vphi}{2}} \cos\frac{\beta}{2}
+ i e^{-i\frac{\vphi}{2}} \sin\frac{\beta}{2}\right)^{\ell-m_1}
\,
\left(e^{-i\frac{\vphi}{2}} \cos\frac{\beta}{2}
+ i e^{i\frac{\vphi}{2}} \sin\frac{\beta}{2}\right)^{\ell+m_1}
e^{im_2\vphi} d\vphi,
\nonumber
\end{eqnarray}
%================================================================
which for the particular case of $\beta=\pi/2$ can be presented
in the following form
%================================================================
\begin{eqnarray}
\label{INT-DFUN1}
d^l_{m_2, m_1}(\frac{\pi}{2}) = (-1)^{\frac{l-m_1}{2}}\frac{2^l}{\pi}
\left[\frac{(l+m_2)!(l-m_2)!}{(l+m_1)!(l-m_1)!}\right]^{\frac12}
\,
\int_0^{\pi}(\sin\alpha)^{l-m_1}(\cos\alpha)^{l+m_1}
e^{2im_2\alpha} d\alpha
\end{eqnarray}
%================================================================
Using now the formulas \cite{BE2}
%================================================================
\begin{eqnarray*}
\cos(2n\alpha)= T_{n}(\cos 2\alpha),
\qquad
\sin(2n\alpha)= \sin 2\alpha \cdot U_{n-1}(\cos 2\alpha),
\end{eqnarray*}
%================================================================
where $T_l(x)$ and $U_l(x)$ are Tchebyshev polynomials of the first and
second kind. After integrating over $\alpha$, we obtain a representation
of the Wigner $d$-function for angles $\pi/2$ in terms of the
hypergeometrical function $_3F_2(1)$
%===============================================================
\begin{eqnarray}
\label{D1}
d^l_{m_2, m_1}(\frac{\pi}{2}) =
\frac{(-1)^{\frac{l-m_1}{2}}}{\sqrt{\pi} l!}
\, \sqrt{(l+m_2)!(l-m_2)!}
\end{eqnarray}
%==============================================================
\begin{eqnarray*}
\times
\left\{\matrix{
\left\{\frac{\Gamma\left(\frac{l+m_1+1}{2}\right)
\Gamma\left(\frac{l-m_1+1}{2}\right)}
{\Gamma\left(\frac{l+m_1}{2}+1\right)\Gamma\left(\frac{l-m_1}{2}+1\right)}
\right\}^{\frac12}
\,
{_3F_2}
\left(\matrix{
- m_2, m_2, \frac{l+m_1+1}{2}
\cr
\frac{1}{2},
l+1 \cr}\Bigg|1\right),
\qquad (l+m_1)- {\rm even},
\cr
\cr
\cr
\cr
\frac{2il}{(l+1)}\,
\left\{\frac{\Gamma\left(\frac{l+m_1}{2}+1\right)
\Gamma\left(\frac{l-m_1}{2}+1\right)}
{\Gamma\left(\frac{l+m_1+1}{2}\right)\Gamma\left(\frac{l-m_1+1}{2}\right)}
\right\}^{\frac12}
\,
{_3F_2}
\left(\matrix{
-m_2+1, m_2+1 , \frac{l+m_1}{2}+1
\cr
\frac{3}{2},
l+2\cr}\Bigg|1\right),
\,\, (l+m_1) - {\rm odd}.
\cr}\right.
\end{eqnarray*}
%================================================================
For large $R$ we put
%================================================================
\begin{eqnarray}
\label{LIM1}
l \sim k R, \quad
m_1 \sim k_1 R, \quad
\theta_1  \sim \frac{r}{R}, \quad
\theta_1' \sim \frac{y}{R}, \quad
\theta_2' \sim \frac{x}{R},
\end{eqnarray}
%================================================================
where $k^2=k_1^2+k_2^2$. Using the asimptotic formulas for $_3F_2(1)$
function (\ref{GAMMA-FUN10}) and $\Gamma$ function (\ref{GAMMA-FUN1}),
we get
%==============================================================
$$
\lim_{R\rightarrow\infty}
(-1)^{-\frac{l-|m_1|}{2}} \,
\sqrt{R}
\,
{d}_{m_2, m_1}^{l}(\frac{\pi}{2})
=
\sqrt{\frac{2}{\pi k}}
$$
\begin{eqnarray}
\label{DW}
\times
\left\{\matrix{
\left(\frac{k^2}{k_2}\right)^{\frac14}
\,\,\,
{_2F_1}\left(- m_2, m_2; \, \frac12; \, \frac{k+k_1}{2k}\right),
\cr
 -i m_2 \,
\left(\frac{k_2}{k^2}\right)^{\frac14}
\,
{_2F_1}
\left(-m_2+1, m_2+1; \,
\frac{3}{2}; \, \frac{k+k_1}{2k} \right),
\cr}\right.
=
(-1)^{\frac{3m_2}{2}}
\sqrt{\frac{2}{\pi k_2}}
\,
\left\{\matrix{
\, \cos m_2 \varphi,
\cr
\cr
i \sin m_2 \varphi,
\cr}\right.
\end{eqnarray}
%===============================================================
with $\cos\varphi = k_1/k$ and for $(\ell+m_1)$ even or odd
respectively.

Multiplying now the interbasis expansion (\ref{EXP1}) by the factor
$(-1)^{-\frac{l-|m_1|}{2}}$, and taking the contraction limit
$R\to\infty$ we obtain ($\theta\equiv \theta_2, \,  m \equiv m_2$)
%===============================================================
\begin{eqnarray}
e^{ik_{1}x}
\left\{\matrix{\cos{k}_{2}y \cr
\sin{k}_{2}y \cr}
\right\}
=
\sum_{m = - \infty}^{\infty}
\,
(i)^{|m|}
\left\{\matrix{\cos{m \varphi}
\cr
 - \sin{m\varphi}\cr}
\right\}
\,
J_{|m|}(kr)
\,
e^{im\theta},
\end{eqnarray}
%===============================================================
or in exponential form
%===============================================================
\begin{eqnarray}
\label{EXP21}
e^{ikr \cos(\theta-\varphi)}
=
\sum_{m = - \infty}^{\infty}
\,
(i)^{m}
\,
J_{m}(kr)
\,
e^{im (\theta-\varphi)}.
\end{eqnarray}
%===============================================================
The inverse expansion is
%===============================================================
\begin{eqnarray}
\label{EXP22}
J_{m} (kr) e^{im\theta}
=
\frac{(-i)^{m}}{2\pi}\,
\int_0^{2\pi}
e^{im\varphi - ikr\cos(\theta-\varphi)}\, d\varphi.
\end{eqnarray}
%===============================================================
For $\theta=0$ the two last formulas are equivalent to the well know
formulas in the theory of Bessel functions \cite{BE2}, namely expansions
of plane waves in terms of cylindrical ones and vice versa.

\subsection{Solutions of the Lam{\'e} equation}

Let us consider eq.~(\ref{LB-HELM2}) on the sphere $S_{2}$ and
separate variables in the elliptic coordinates (\ref{ALGEB1}).
We obtain two ordinary differential equation of the form
%===============================================================
\begin{eqnarray}
\label{LAME-0}
\frac{d^{2}\psi}{d\rho^{2}}+\frac{1}{2}
\left\{\frac{1}{\rho-a_{1}}+
\frac{1}{\rho-a_{2}}+\frac{1}{\rho-a_{3}}
\right\}
\frac{d\psi}{d\rho}
+ \frac{1}{4}
\left\{\frac{\lambda-l(l+1)\rho}{(\rho-a_{1})(\rho-a_{2})
(\rho-a_{3})}\right\} \psi =0
\end{eqnarray}
%===============================================================
or equivalently
%===============================================================
\begin{eqnarray}
\label{LAME-1}
4\sqrt{P(\rho)}\frac{d}{d\rho} \sqrt{P(\rho)}\frac{d\psi}{d\rho}-
\{l(l+1)\rho-\lambda\}\psi=0
\end{eqnarray}
%===============================================================
where
%===============================================================
$$
P(\rho)=(\rho-a_{1})(\rho-a_{2})(\rho-a_{3}).
$$
%===============================================================
Eq.(\ref{LAME-0}) is the Lam{\'e} equation in algebraic form.
It is a Fuchsian type equation with 4 regular singularities
(at $a_1$, $a_2$, $a_3$ and $\infty$)
\cite{PAWI,LS1,LUKA1,WITWAT,HOB,ARSS}.

Its general solution can be represented by a series expansion about
any one of the singular points $a_k$ as
%===============================================================
\begin{eqnarray}
\label{LAME-2}
\psi(\rho)
=(\rho-a_1)^{\alpha_1/2}(\rho-a_2)^{\alpha_2/2}
(\rho-a_3)^{\alpha_3/2}
\sum_{t=0}^{\infty}b_{t}^{(k)}(\rho-a_k)^t,
\end{eqnarray}
%===============================================================
where we have
$$
\alpha_{j}(\alpha_{j}-1)=0, \,\,\,\, j = 1,2,3
$$
and can choose $k$ equal to 1,2, or 3.

Substituting (\ref{LAME-2}) into the Lam{\'e} equation (\ref{LAME-0})
we obtain a three term recursion relation for $b_{t}^{k}$
%===============================================================
\begin{eqnarray}
\label{LAME-3}
\beta_{t}^{(k)}b_{t+1}^{(k)}+
[\gamma_{t}^{(k)} + \lambda - l(l+1) a_{k}] b_{t}^{(k)}
+
(2t+\alpha-l-2)(2t+\alpha+l-1)b_{t-1}^{(k)}
= 0
\end{eqnarray}
%===============================================================
with
%===============================================================
$$
\alpha
= \alpha_1+\alpha_2+\alpha_3,
\,\,\,\,
\alpha_{ik}=\alpha_i-\alpha_k,
\,\,\,\,
b_{-1}=0
$$
%===============================================================
%===============================================================
\begin{eqnarray}
\beta_{t}^{(k)}&=&4(a_i - a_k)(a_j - a_k)(t+1)(t+\alpha_k+1/2)
\,\,\,\,\,
(i,j,k \,\,{\rm cyclic})
\nonumber\\[2mm]
\gamma_{t}^{(k)}
&=&
- (a_i - a_k)(2t+\alpha_k+\alpha_j)^2 - (a_j - a_k)
(2t+\alpha_k+\alpha_i)^2.
\end{eqnarray}
%===============================================================
The expansion (\ref{LAME-2}) represents a Lam{\'e} function. Since
we are interested in representations of O(3), the sum in $\psi(\rho)$
must be a polynomial of order $N$, i.e. we must have
%===============================================================
\begin{eqnarray}
\label{LAME-4}
b_N \ne 0,\,\,\,\, b_{N+1} = b_{N+2} = \cdots =0
\end{eqnarray}
%===============================================================
for some $N$. The condition for this is that we have
%===============================================================
\begin{eqnarray}
\label{LAME-5}
l=2N+\alpha
\end{eqnarray}
%===============================================================
and we obtain a secular equation for the eigenvalues $\lambda$,
i.e. the separation constant in elliptic coordinates, by requiring
that the determinant of the homogeneous linear system (3.14) for
\{$b_0,b_1,\cdots, b_N$\} should vanish. Since $N$ and $l$ must be
integers, eq.~(\ref{LAME-5}) implies that $\alpha$ and $l$ must have
the same parity.

Numerous further properties of the Lam{\'e} polynomials, in the
context of representations of the group O(3), in the O(3)$\supset$D$_2$
basis, were established e.g. in Ref. \cite{PAWI,LS1,LUKA1}.

Here let us just represent the basis functions as
%===============================================================
\begin{eqnarray}
\label{LAME-6}
\Psi^{pq}_{l\lambda}(\rho_1, \rho_2) = A^{pq}_{l\lambda}
\psi^{pq}_{l\lambda}(\rho_1)
\psi^{pq}_{l\lambda}(\rho_2),
\end{eqnarray}
%===============================================================
where $A^{pq}_{l\lambda}$ is some normalization constant.
The labels $p,q$ take values $\pm 1$ and identify representations
of D$_2$. For each value of $l$ the values of $p,q$ and
$\lambda$  label $(2l+1)$ different states. Since a given
representations $(p, q)$ of D$_2$ can figure more than once
in the reduction of a representation of O(3) corresponding to
a given $l$, we are faced with a "missing label problem",
resolved by the quantum number $\lambda$, i.e. the operator
$Q$ of eq.~(\ref{ENALGEB1}).

The expansions that we shall use for the Lam{\'e} polynomials in
(\ref{LAME-6}) are as in eq.~(\ref{LAME-2}), but the summation over
$t$ is from $t=0$ to $t=N$.

\subsection{Elliptic basis on $S_2$ to Cartesian basis on $E_2$}

We choose elliptic coordinates on $S_2$ as in eq.~(\ref{ALGEB1}), but
with $a_1<a_3<a_2$, as in eq.~(\ref{CART-ELLIP0}). We write the basis
functions as in eq.~(\ref{LAME-6}) with
%===============================================================
\begin{eqnarray}
\label{LAME-7}
\psi_{l\lambda}(\rho_{1})=(\rho_{1}-a_1)^{\alpha_{1}/2}
(\rho_{1}-a_2)^{\alpha_{2}/2}(\rho_{1}-a_3)^{\alpha_{3}/2}
\sum_{t=0}^{N}b_{t}^{(1)}(\rho_{1}-a_{1})^{t}
\nonumber\\[2mm]
\psi_{l\lambda}(\rho_{2})=(\rho_{2}-a_1)^{\alpha_{1}/2}
(\rho_{2}-a_2)^{\alpha_{2}/2}(\rho_{2}-a_3)^{\alpha_{3}/2}
\sum_{t=0}^{N}b_{t}^{(2)}(\rho_{2}-a_{2})^{t}
\end{eqnarray}
%===============================================================
as in eq.~(\ref{LAME-2}). The coefficients $b_{t}^{(j)} \,(j=1,2)$
satisfy the recursion relation (\ref{LAME-3}) and we have $N=(l-\alpha)/2$.
We use the coordinates $\xi_{1}$ and $\xi_{2}$ introduced in
eq.~(\ref{CART-ELLIP1}) (for $a\equiv a_{3}-a_{1}=a_{2}-a_{3}$).
Eq.~(\ref{LAME-8}) reduces to
%===============================================================
\begin{eqnarray}
\label{LAME-8}
\psi_{l\lambda}(\xi_{1})
&=& (-1)^{\frac{\alpha_{2}+\alpha_{3}}{2}}a^{\frac{\alpha}{2}}
(1-\xi_{1})^{\frac{\alpha_{1}}{2}}(1+\xi_{1})^{\frac{\alpha_{2}}{2}}
\xi_{1}^{\frac{\alpha_{3}}{2}}\sum_{t=0}^{N}C_{t}^{1}(1-\xi_{1})^{t}
\nonumber\\[2mm]
\psi_{l\lambda}(\xi_{2})
&=&(-1)^{\frac{\alpha_{2}}{2}}a^{\frac{\alpha}{2}}
(1-\xi_{2})^{\frac{\alpha_{2}}{2}}(1+\xi_{2})^{\frac{\alpha_{1}}{2}}
\xi_{2}^{\frac{\alpha_{3}}{2}}\sum_{t=0}^{N}C_{t}^{2}(1-\xi_{2})^{t}
\end{eqnarray}
%===============================================================
with $C_{t}^{(1)}=a^{t}b_{t}, \, C_{t}^{(2)}=(-a)^{t}b_{t}$.
The recursion relations (\ref{LAME-3}) now imply
%===============================================================
\begin{eqnarray}
\label{LAME-9}
8(t+1)(t+\alpha_{1}+\frac{1}{2})C_{t+1}^{(1)}+\{\mu ^{(1)}-2(2t+\alpha_{1}+
\alpha_{3})^{2}-(2t+\alpha_{1}+\alpha_{2})^{2}\}C_{t}^{(1)}
\nonumber\\[2mm]
+
(2t+\alpha-l-2)(2t+\alpha+l-1)C_{t-1}^{(1)}=0
\nonumber\\
- 8(t+1)(t+\alpha_{2}
+\frac{1}{2})C_{t+1}^{(2)}+\{\mu ^{(2)}+2(2t+\alpha_{2}+
\alpha_{3})^{2}+(2t+\alpha_{1}+\alpha_{2})^{2}\}C_{t}^{(2)}
\\[2mm]-
(2t+\alpha-l-2)(2t+\alpha+l-1)C_{t-1}^{(2)}=0
\nonumber
\end{eqnarray}
%===============================================================
where
%===============================================================
\begin{eqnarray}
\label{LAME-10}
\mu ^{(j)} = \frac{1}{a}[\lambda-a_{j}l(l+1)], \,\,\,\,
j=1,2
\end{eqnarray}
%===============================================================

The contraction limit is taken using eq.~(\ref{CART-ELLIP1}) to relate
$\xi_{1,2}$ to the Cartesian coordinates on $E_2$. Taking $l \sim kR$
we find
%===============================================================
\begin{eqnarray}
\label{LAME-11}
\mu ^{(1)}\rightarrow2R^{2}k_{1}^{2},
\,\,\,\,\,\,\,\,
\mu ^{(2)}\rightarrow-2R^{2}k_{2}^{2},
\,\,\,\,\,\,\,\,\,
k= \sqrt{k_1^2 + k_2^2}.
\end{eqnarray}
%===============================================================
For $R\rightarrow\infty$ the recursion relations (\ref{LAME-9})
simplify to two term ones that can be solved to obtain
%===============================================================
\begin{eqnarray}
\label{LAME-12}
C_{t}^{(j)} =
\frac{R^{2t}}{(\alpha_{j}+\frac{1}{2})_{t}}\cdot
\left(\frac{-k_{j}^{2}}{4}\right)^{t}\cdot
\frac{1}{t!}
\end{eqnarray}
%===============================================================
with
%===============================================================
$$
\label{LAME-13}
\left(\alpha_{j}+\frac{1}{2}\right)_{t}
= \left(\alpha_{j}+\frac{1}{2}\right)
\left(\alpha_{j}+\frac{3}{2}\right)
\cdots
\left(\alpha_{j}-\frac{3}{2}+t\right),
\,\,\,\, t\geq1
\,\,\,\,
\left(\alpha_{j}+\frac{1}{2}\right)_{0}=1.
$$
%===============================================================
Substituting (\ref{LAME-13}) into (\ref{LAME-8}) we obtain
%===============================================================
\begin{eqnarray}
\label{LAME-14}
\psi_{l\lambda}(\xi_{1})
&=&
(-1)^{\frac{\alpha_{2}+\alpha_{3}}{2}}
\frac{a^{\frac{\alpha}{2}}}{R^{\alpha_{1}}}x^{\alpha_{1}}
{_{0}F_{1}}\left(
\alpha_{1}+\frac{1}{2};-\frac{k_{1}^{2}x^{2}}{4}\right)
\nonumber\\[2mm]
\psi_{l\lambda}(\xi_{2})
&=&(-1)^{\frac{\alpha_{2}}{2}}
\frac{a^{\frac{\alpha}{2}}}{R^{\alpha_{2}}}y^{\alpha_{2}}
{_{0}F_{1}}\left(\alpha_{2}+\frac{1}{2};
- \frac{k_{2}^{2}y^{2}}{4}\right).
\end{eqnarray}
%===============================================================
Using now the formula (\ref{SPHER-CARTES0}) we find the  contraction
limit:
%===============================================================
\begin{eqnarray}
\label{LAME-15}
A_{l\lambda}^{pq}(R)\psi_{l\lambda}(\xi_{1},\xi_{2})
&\rightarrow&
A_{l\lambda}^{pq}(R)\psi_{k_{1}}(x)\psi_{k_{2}}(y)
\nonumber
\\[4mm]
=
A^{p q}_{l \lambda}
(-1)^{\frac{\alpha_{3}}{2}}a^{\frac{\alpha}{2}}
&\times&
\cases{\cos{k}_{1}x \cos{k}_{2}y,
&$\alpha_{1}=0, \alpha_{2}=0$
\cr
\cr
- \frac{1}{k_{2}R}\cos{k}_{1}x \sin{k}_{2}y,
&$\alpha_{1}=0, \alpha_{2}=1$
\cr
\cr
- \frac{1}{k_{1}R}\sin{k}_{1}x \cos{k}_{2}y,
&$\alpha_{1}=1, \alpha_{2}=0$
\cr
\cr
-\frac{1}{k_{1}k_{2}R^2}\sin{k}_{1}x \sin{k}_{2}y,
&$\alpha_{1}=1, \alpha_{2}=1$.
\cr
}
\end{eqnarray}
%===============================================================

\subsection{Elliptic basis on $S_2$ to elliptic basis on $E_2$}

Let us start from the elliptic coordinates (\ref{ALGEB1}) with
$a_1\leq \rho_1 \leq a_2 \leq \rho_2 \leq a_3$. We take the limit
$R\rightarrow \infty$, $a_3\rightarrow \infty$ with $\sqrt{a_3}/R$,
$a_1$ and $a_2$ finite. We introduce a constant $D$ as in
eq.~(\ref{ELLIP1}). Elliptic coordinates on the plane $E_2$ are
introduced via eq.~(\ref{ELLIP-TRIG1}), so that the Cartesian coordinates
$(x,y)$ are expressed in terms of the elliptic ones $(\xi, \eta)$
as in eq.~(\ref{ELLTRIG1}). Let us first take the limit in the separated
equations (\ref{LAME-0}). Going over to the variables $(\xi, \eta)$ from
$(\rho_1, \rho_2)$ we obtain for $R\rightarrow \infty$:
%===============================================================
\begin{eqnarray}
\label{LAME-16}
\frac{d^2\psi_1}{d\eta^2} +
\left\{\mu - \frac{k^2D^2}{2}
\left(\frac{a_2+a_1}{a_2-a_1}\right)
- \frac{k^2D^2}{2}\cos 2\eta
\right\}\psi_1=0
\end{eqnarray}
%===============================================================
%===============================================================
\begin{eqnarray}
\label{LAME-17}
\frac{d^2\psi_2}{d\xi^2} +
\left\{\mu - \frac{k^2D^2}{2}
\left(\frac{a_2+a_1}{a_2-a_1}\right)
- \frac{k^2D^2}{2}\cosh 2\xi
\right\}\psi_2=0
\end{eqnarray}
%===============================================================
with
%===============================================================
$$
\mu = \frac{\lambda}{a_3},
\,\,\,\,\,\,\,\,\,
l \sim k R.
$$
%===============================================================

In (\ref{LAME-16}) we recognize the standard form of the Mathieu
equation, whereas eq.~(\ref{LAME-17}) is a modified Mathieu equation
\cite{BE3}. Thus, in the contraction limit, Lam{\'e} functions will
go over into Mathieu ones. Moreover, periodic solutions of the Lam{\'e}
equation go over into periodic solutions of eq.~(\ref{LAME-16}).

The contraction limit can also be taken directly in the Lam{\'e}
polynomials, using the expansion (\ref{LAME-2}) (cut off at $t=N$).
The result that we obtain is
%===============================================================
\begin{eqnarray}
\lim_{R\rightarrow\infty}
\frac{\Psi_{l \lambda}(\rho_1)}{R^{\alpha_3}}
=(a_2-a_1)^{\frac{\alpha}{2}}
\frac{(-1)^{\frac{\alpha_2+\alpha_3}{2}}}{D^{\alpha_3}}
(\cos\eta)^{\alpha_1}(\sin\eta)^{\alpha_2}
\sum_{t=0}^{\infty}C_{t}(\cos\eta)^{2t},
\end{eqnarray}
%===============================================================
%===============================================================
\begin{eqnarray}
\lim_{R\rightarrow\infty}
\frac{\Psi_{l \lambda}(\rho_2)}{R^{\alpha_3}}
=(a_2-a_1)^{\frac{\alpha}{2}}
\frac{(-1)^{\frac{\alpha_3}{2}}}{D^{\alpha_3}}
(\cosh\xi)^{\alpha_1}(\sinh\xi)^{\alpha_2}
\sum_{t=0}^{\infty}C_{t}(\cosh\xi)^{2t},
\end{eqnarray}
%===============================================================
where the expansion coefficients $C_t$ satisfy recursion
relations obtained from eq.~(\ref{LAME-3}), namely
%===============================================================
\begin{eqnarray}
4(t+1)(t+ 1/2+ \alpha_1) C_t +
\{\mu - (2t+\alpha_1 + \alpha_2)^2\} C_t -
k^2 D^2 C_t = 0.
\end{eqnarray}
%===============================================================

\subsection{Elliptic basis on $S_2$ to parabolic basis on $E_2$}

Let us consider the contraction limit for the Lam{\'e} equations
(\ref{LAME-0}). To do this we use equations (\ref{EL-SPHER11}) with
$a_3 - a_2 = a_2 - a_1 = a$ i.e. $k=k'=1/\sqrt{2}$, together
with eq.~(\ref{ELLIM000}), to obtain
%===============================================================
\begin{eqnarray}
\rho_1 \sim a_1 + a \left(-1 + \frac{u^2}{2R}\right),
\,\,\,\,\,\,
\rho_2 \sim a_1 + a \left(1 + \frac{v^2}{R}\right)
\end{eqnarray}
%===============================================================
The equation (\ref{LAME-0}) for $\rho = \rho_1$ and $\rho = \rho_2$
in the limit $R\rightarrow\infty$, with $l^2 \sim k^2R^2$
and {$\lambda - a_2 l(l+1) = \mu R a$}, yields the two equations
%===============================================================
\begin{eqnarray}
\label{LAME-18}
\frac{d^2\psi_1}{du^2} +
(k^2 u^2 + \mu )\psi_1 = 0,
\qquad
\frac{d^2\psi_2}{dv^2} +
(k^2 v^2 - \mu )\psi_2 = 0
\end{eqnarray}
%===============================================================
respectively.

Thus the Lam{\'e} equations in the contraction limit go over into
the equations (\ref{LAME-18}) for parabolic cylinder functions
\cite{BE2}. The same is of course true for solutions.
The expansion (\ref{LAME-2}) is not suitable for the contraction
limit. In view of eq.~(\ref{ELLIM000}) we need expansions in terms
of the variables $(1+\sn\alpha)$ and $(1-\sqrt{2}\dn\beta)$. This
is not hard to do, following for instance methods used in Ref.\cite{MPS}
to relate the wave functions of a two-dimensional hydrogen atom,
calculated in different coordinate systems. The formulas are cumbersome,
so we shall not present them here.

\subsection{Contractions of basis functions from $H_2$ to $E_2$ and
$E_{1,1}$}

{\bf 5.6.1. Pseudo-spherical basis on $H_2$ to polar basis on $E_2$}.

The pseudo-spherical eigenfunctions $\Psi_{\rho m}(\tau, \varphi)$
normalized to the Dirac delta-function, have the form:
%==============================================================
\begin{eqnarray}
\label{LEGENDRE1}
\Psi_{\rho m}(\tau,\varphi)
= \sqrt{\frac{\rho\sinh\pi\rho}
{2\pi^2 R}}\mid\Gamma(\frac{1}{2}+i\rho+|m|)\mid \cdot
P_{i\rho-\frac12}^{|m|} (\cosh\tau)\exp(im\varphi),
\end{eqnarray}
%==============================================================
where $m = 0,\pm 1,\pm 2,...$. In the contraction limit
$R\rightarrow\infty$ we put: $\tanh\tau \sim \tau \sim r/R,\,\rho \sim kR$.
Rewriting the Legendre function in terms of hypergeometric function as
\cite{BE1}
%==============================================================
$$
P_{i\rho-\frac12}^{|m|}(\cosh\tau)
=
\frac{\Gamma(\frac{1}{2}+i\rho+|m|)}{\Gamma(\frac{1}{2}+i\rho-|m|)}
\,
\frac{1}{|m|! 2^{|m|}}
\,
{_2F_1}\left(\frac{1}{2} + |m| + i\rho,
\frac{1}{2} + |m| - i\rho; 1 + |m|; - \sinh^2\frac{\tau}{2}\right).
$$
%==============================================================
Then using the asymptotic formula for hypergeometrical function
$_2F_1$ and $\Gamma$ function
%===============================================================
\begin{eqnarray}
\label{GAMMA1}
\lim_{\mid y \mid \rightarrow\infty}\mid\Gamma(x+iy)\mid
 \exp(\frac{\pi}{2}\mid y\mid)\mid y\mid ^{\frac{1}{2}-x} =
 \sqrt{2\pi}
\end{eqnarray}
%===============================================================
we obtain in the contraction limit $R\rightarrow\infty$:
%==============================================================
\begin{eqnarray*}
\lim_{R\rightarrow\infty}\Psi_{\rho m} (\tau,\varphi) =
\sqrt{k}\cdot J_{|m|}(kr)
\cdot\frac{e^{im\varphi}}{\sqrt{2\pi}},
\end{eqnarray*}
%================================================================

\vspace{0.3cm}
\noindent
{\bf 5.6.2. Pseudo-spherical basis on $H_2$ to Cartesian basis on
$E_{1,1}$}.

Taking the Legendre function in eq.~(\ref{LEGENDRE1}) in terms of
two hypergeometric functions \cite{BE1}
%==============================================================
\begin{eqnarray*}
P_{i\rho-1/2}^{m}(\cosh\tau) =
\frac{\sqrt{\pi} 2^{m}
(\sinh\tau)^{-m}}{\Gamma(\frac{3}{4}-\frac{m+i\rho}{2})
\Gamma(\frac{3}{4}-\frac{m-i\rho}{2})}
\,
\Biggl\{
2\cosh\tau \frac{\Gamma(\frac{3}{4}-\frac{m+i\rho}{2})
\Gamma(\frac{3}{4}-\frac{m-i\rho}{2})}
{\Gamma(\frac{1}{4}-\frac{m+i\rho}{2})
\Gamma(\frac{1}{4}-\frac{m-i\rho}{2})}
\\[2mm]
{_2 F_1}\left(\frac{3}{4}-\frac{m+i\rho}{2},
\frac{3}{4}-\frac{m-i\rho}{2}; \frac{3}{2};
\cosh^{2}\tau\right)
+
{_2 F_1}\left(\frac{1}{4}-\frac{m+i\rho}{2},
\frac{1}{4}-\frac{m-i\rho}{2}; \frac{1}{2};
\cosh^{2}\tau\right)
\Biggr\}.
\end{eqnarray*}
%=============================================================
Putting for large $R$
%=============================================================
$$
\rho \sim kR,
\,\,
m \sim k_1 R,
\,\,\,\,
\coth\tau \sim \frac{t}{R},
\,\,\,\,
\cot\varphi \sim \frac{x}{R},
;
\,\,\,\,
k^2 + k_1^2 = k_0^2.
$$
%==============================================================
Using two asymptotic formulas
%==============================================================
\begin{eqnarray*}
\lim_{R\rightarrow\infty}
{_2 F_1}\left(\frac{1}{4}-\frac{m+i\rho}{2},
\frac{1}{4}-\frac{m-i\rho}{2}; \frac{1}{2}; \cosh^{2}\tau\right)
= {_0 F_1}\left(\frac{1}{2};
- \frac{k_0^2 t^2}{4}\right) = \cos(k_0t),
\end{eqnarray*}
%=============================================================
\begin{eqnarray*}
\lim_{R\rightarrow\infty}
{_2 F_1}\left(\frac{3}{4}-\frac{m+i\rho}{2},
\frac{3}{4}-\frac{m-i\rho}{2}; \frac{3}{2}; \cosh^{2}\tau\right)
= {_0F_1}\left(\frac{3}{2}; -\frac{k_0^2 t^2}{4}\right)
= \frac{\sin(k_0t)}{k_0t}.
\end{eqnarray*}
%=============================================================
and formula (\ref{GAMMA-FUN1}) we finally obtain
%=============================================================
\begin{eqnarray}
\lim_{R\rightarrow\infty} \sqrt{R} |\Gamma(i\rho)|
\Psi_{\rho m}(\tau,\varphi) =
\sqrt{\frac{2}{k_0}} e^{ik_0t - ik_1x}.
\end{eqnarray}
%=============================================================

\subsection{Contractions for equidistant basis on $H_2$}

{\bf 5.7.1. Equidistant basis on $H_2$ to Cartesian basis on $E_2$}.

In the equidistant system the normalized eigenfunctions $\Psi_{\rho\lambda}
(\tau_1,\tau_2)$ have the form:
%===============================================================
\begin{eqnarray*}
  \Psi_{\rho\lambda}(\tau_1,\tau_2) =
  \sqrt{\frac{\rho\sinh\pi\rho}{\cosh^2\pi\lambda+\sinh^2\pi\rho}}
  \cdot (\cosh\tau_1)^{-1/2}
  P_{i\lambda-1/2}^{i\rho}(-\tanh\tau_1)\cdot e^{i\lambda\tau_2}.
\end{eqnarray*}
%===============================================================
To perform the contraction we write the Legendre function in terms of
hypergeometric function \cite{BE1}
%===============================================================
\begin{eqnarray*}
 P^{i\rho}_{i\lambda-1/2}(-\tanh\tau_1) = \frac{\sqrt{\pi}2^{i\rho}
 (\cosh\tau_1)^{-i\rho}}
 {\Gamma\left(\frac{3}{4} - a \right)
 \Gamma\left(\frac{3}{4} - b \right)}
 \Biggl\{ {_2F_1}\left(\frac{1}{4} + a,
 \frac{1}{4} + b; \frac{1}{2}; \tanh^2\tau_1\right)
\\[2mm]
+ 2 \tanh\tau_1 \frac{\Gamma\left(\frac{3}{4} - a \right)
 \Gamma\left(\frac{3}{4} - b \right)}
 {\Gamma\left(\frac{1}{4} - a \right)
 \Gamma\left(\frac{1}{4} - b \right)}
 {_2F_1}\left(\frac{3}{4} - a,
 \frac{3}{4} - b; \frac{3}{2}; \tanh^2\tau_1\right)
 \Biggr\},
\end{eqnarray*}
%===============================================================
where $a = i(\rho-\lambda)/2$, $b = i(\rho+\lambda)/2$.
For large $R$ we put $\rho\sim kR,\,\,\lambda\sim k_1R;\,\,\,\,  \tau_2
 \sim x/R,\,\,\,   \tau_1\sim y/R$
where $x,y$ are the Cartesian coordinates. Then using the asymptotic
formulas:
%===============================================================
\begin{eqnarray*}
\lim_{R\rightarrow\infty}
{_2F_1}\left(\frac{1}{4} + a, \frac{1}{4} + b; \frac{1}{2};
\tanh^2\tau_1\right)
&=& {_0F_1} \left(\frac{1}{2}; - \frac{y^2k_2^2}{4}\right)
= \cos k_2 y,
\\[2mm]
\lim_{R\rightarrow\infty}
{_2F_1}\left(\frac{3}{4} - a, \frac{3}{4} - b; \frac{3}{2};
\tanh^2\tau_1\right)
&=& {_0F_1} \left(\frac{3}{2}; - \frac{y^2k_2^2}{4}\right)
= \frac{1}{k_2 y} \sin k_2 y,
\end{eqnarray*}
%===============================================================
where $k_1^2+k_2^2=k^2$, we finally get
%===============================================================
\begin{eqnarray*}
 \lim_{R\rightarrow\infty} \Psi_{\rho\lambda}(\tau_1,\tau_2) =
 \sqrt{\frac{k}{\pi k_2}} \exp(ik_1x + ik_2y).
\end{eqnarray*}
%===============================================================

\vspace{0.3cm}
\noindent
{\bf 5.7.2. Contraction from equidistant basis on  $H_2$ to polar on
$E_{1,1}$}.

Writting the Legendre function in terms of hypergeometric functions
\cite{BE1}
%==============================================================
\begin{eqnarray*}
P_{i\lambda-1/2}^{i\rho}(\tanh\tau_1) =
\frac{1}{\sqrt{2\pi}}(\sinh\tau_1)^{i\rho}
\cdot
\Biggl\{
2^{-i\lambda}(\coth\tau_1)^{i\lambda+1/2}
\,
\frac{\Gamma(-i\lambda)}{\Gamma(\frac{1}{2}-
i(\rho+\lambda))}
\\[2mm]
{_2 F_1}\left(\frac{1}{4}-\frac{i(\rho-\lambda)}{2},
\frac{3}{4}-\frac{i(\rho-\lambda)}{2}; \, 1+i\lambda; \,
\coth^{2}\tau_1
\right) +
2^{i\lambda}(\coth\tau_1)^{-i\lambda+1/2}
\\[2mm]
\frac{\Gamma(i\lambda)}{\Gamma(\frac{1}{2}-i(\rho-\lambda)}
{_2 F_1}\left(\frac{1}{4}-\frac{i(\rho+\lambda)}{2},
\frac{3}{4}-\frac{i(\rho+\lambda)}{2}; \, 1-i\lambda;
\, \coth^{2}\tau_1 \right)
\Biggr\}.
\end{eqnarray*}
%=============================================================
Putting for large $R$: $\rho \sim kR$ and $\cosh\tau_1 \sim
\frac{r}{R}$, and using the asymptotic formulas for hypergeometric
functions
%==============================================================
$$
\lim_{R\rightarrow\infty}
{_2 F_1}\left(\frac{1}{4}-\frac{i(\rho-\lambda)}{2},
\frac{3}{4}-\frac{i(\rho-\lambda)}{2}; \, 1+i\lambda; \,
\coth^{2}\tau_1 \right)
=
\Gamma(1+i\lambda)\left(\frac{kr}{2}\right)^{-i\lambda}
J_{i\lambda}(kr),
$$
$$
\lim_{R\rightarrow\infty}
{_2 F_1}\left(\frac{1}{4}-\frac{i(\rho+\lambda)}{2},
\frac{3}{4}-\frac{i(\rho+\lambda)}{2}; \, 1-i\lambda;
\, \coth^{2}\tau_2 \right)
=
\Gamma(1-i\lambda)\left(\frac{kr}{2}\right)^{+i\lambda}
J_{-i\lambda}(kr),
$$
%=============================================================
we obtain
%==============================================================
\begin{eqnarray*}
\lim_{R\rightarrow\infty}\frac{1}{\sqrt{R}}\Psi_{\rho\lambda}
(\alpha,\tau_2) = \sqrt{\frac{k}{2}}
H_{i\lambda}^{(1)}(kr) e^{i\lambda (\tau_2 + i\frac{\pi}{2})},
\end{eqnarray*}
%=============================================================
where $H_{\nu}^{(1)}(z)$ is the first kind of Hankel function.

%-----------------------------------------------------------------
\section*{Acknowledgements}
G.P. thank the support of the Direcci\'on General de Asuntos del
Personal Aca\-d\'e\-mico, Universidad Nacional Aut\'onoma de M\'exico
({\sc dgapa--unam}) by the grant IN112300 {\it Optica Matem\'atica}
and acknowledges the Consejo Nacional de Ciencia y Tecnolog\'{\i}a
(M\'exico) for a C\'atedra Patrimonial Nivel II.
The research od P.W. is partly supported by research grants from NSERC
of Canada and FCAR du Quebec. G.P. and P.W. benefitted from a NATO
collaborative linkage grant which made mutual visits possible.

\end{document}